\documentclass[journal]{IEEEtran}

\usepackage{amsmath}
\usepackage{amssymb}
\usepackage[inline,shortlabels]{enumitem}
\usepackage[normalem]{ulem}
\usepackage{comment}
\usepackage{xcolor}
\usepackage{mathtools}
\usepackage[numbers]{natbib}
\usepackage{caption}
\usepackage{adjustbox}
\usepackage{multicol}
\usepackage{lipsum}
\usepackage{titlesec}
\usepackage{algorithm,algorithmic}
\usepackage{flushend}
\usepackage{subfigure}
\usepackage{xfrac}
\usepackage{hyperref}

\def\f{$f$}

\def\NN{NN}
\def\RNN{RNN}
\def\LSTM{LSTM}
\def\PyNN{PhyNN}
\def\PyLSTM{PhyLSTM}
\def\LSTM{LSTM}

\def\MSE{MSE}
\def\RNN{RNN}
\def\SoC{SoC}
\def\RoC{RoC}
\def\ECS{ECS}
\def\BasinTemp{$T_{b}$}
\def\tslot{$\Delta t$}
\def\BasinTempNoise{$T^n_{b}$}
\def\ProcessTemp{$T_{p}$}
\def\Qp{$Q_{p}$}
\def\Qt{$Q_{t}$}
\def\cw{$c_{w}$}
\def\Vw{$V_{w}$}
\def\FanPower{$P_{f}$}
\def\timeslot{$t_{n}$}
\def\rhow{$\rho_{w}$}
\def\rhoa{$\rho_{a}$}
\def\madot{$\Dot{m}_a$}
\def\mwdot{$\Dot{m}_w$}
\def\Henter{$H_{a,e}$}
\def\Hleave{$H_{a,l}$}
\def\Weathertuple{$W$}
\newcommand{\etc}{etc.\ }
\newcommand{\cf}{cf.\ }
\newcommand{\eg}{e.g., }

\newcommand{\ie}{i.e., }

\def \figurename {Fig.}
\def \figurenamelong {Figure}  
\def \Sectionname {Section}
\def \Tablename {Table}
\def \Eqname {Eq.}

\def \algoname {Algorithm}
\newcommand{\fref}[1]{\figurename~\ref{#1}}
\newcommand{\Fref}[1]{\figurenamelong~\ref{#1}}
\newcommand{\tref}[1]{\Tablename~\ref{#1}}
\newcommand{\sref}[1]{\Sectionname~\ref{#1}}

\newcommand{\eref}[1]{\Eqname~(\ref{#1})}
\newcommand{\algoref}[1]{\algoname~\ref{#1}}
\newcommand{\qref}[1]{\ref{#1}}

\newcommand{\newtext}[1]{\textcolor{blue}{#1}}
\DeclarePairedDelimiter\floor{\lfloor}{\rfloor}
\begin{document}

\title{Physics Informed LSTM Network for Flexibility Identification in Evaporative Cooling Systems} 

\author{Manu~Lahariya,~\IEEEmembership{}
        Farzaneh~Karami,~\IEEEmembership{}
        Chris~Develder~\IEEEmembership{}
        and~Guillaume~Crevecoeur~\IEEEmembership{} 
        
\thanks{M.\ Lahariya and C.\ Develder are with IDLab, Ghent University - imec, Technologiepark-Zwijnaarde 126, 9052~Ghent, Belgium, e-mail:$\{$manu.lahariya, Chris.Develder$\}$@ugent.be.}
\thanks{F.\ Karami is with Dept.\ of Electromechanical, Systems and Metal Engineering, Ghent University, Core lab EEDT-DC, Flanders Make, Ghent, Belgium and Dept.\  of Computer Science, CODeS, KU Leuven, Gebroeders De Smetstraat 1, 9000 Gent, Belgium,  e-mail: karami.farzaneh@ugent.be}
\thanks{G.\ Crevecoeur is with Dept.\ of Electromechanical, Systems and Metal Engineering, Ghent University, Core lab EEDT-DC, Flanders Make, Ghent, Belgium, e-mail: Guillaume.Crevecoeur@UGent.be}
}

\makeatletter
\def\ps@IEEEtitlepagestyle{%
  \def\@oddfoot{\mycopyrightnotice}%
  \def\@evenfoot{}%
}
\def\mycopyrightnotice{%
  {\footnotesize Published in IEEE Transactions on Industrial Informatics. Copyright~\copyright~2022 IEEE. \\ doi: \url{https://doi.org/10.1109/TII.2022.3173897}}
}

\maketitle

\begin{abstract}
In energy intensive industrial systems, an evaporative cooling process may introduce operational flexibility. Such flexibility refers to a system's ability to deviate from its scheduled energy consumption.  Identifying the flexibility, and therefore, designing control that ensures efficient and reliable operation presents a great challenge due to the inherently complex dynamics of industrial systems. Recently, machine learning models have attracted attention for identifying flexibility, due to their ability to model complex nonlinear behavior. This research presents machine learning based methods that integrate system dynamics into the machine learning models (\eg Neural Networks) for better adherence to physical constraints. We define and evaluate physics informed long-short term memory networks ({\PyLSTM}) and physics informed neural networks ({\PyNN}) for the identification of flexibility in the evaporative cooling process. These physics informed networks approximate the time-dependent relationship between control input and system response while enforcing the dynamics of the process in the neural network architecture. Our proposed {\PyLSTM} provides less than 2\% system response estimation error, converges in less than half iterations compared to a baseline Neural Network ({\NN}), and accurately estimates the defined flexibility metrics. We include a detailed analysis of the impact of training data size on the performance and optimization of our proposed models.
\end{abstract}

\IEEEpeerreviewmaketitle

\section{Introduction}
\label{sec:introduction}
\IEEEPARstart{E}NERGY intensive processes require effective control design for reliable and efficient operation~\cite{Metzger_2011_agentbasedcontrolreview}. An Evaporative Cooling System ({\ECS}) is one such energy intensive process where control can be used to mitigate energy fluctuations. Designing control for {\ECS} requires modeling the complex behavior of a non-linear system to identify its operational flexibility.

Operational flexibility of a system is the ability to deviate from the original planning in terms of energy consumption. Quantification and measurement of operational flexibility remain challenging, both theoretically and practically. This is performed using system identification, which results in parametric models with the capability of predicting the effect of control settings on the systems' dynamic behavior~\cite{wang_2017_opflex}. System identification in energy intensive systems is thus an essential step that assists in estimating operational flexibility, developing control, making pricing/cost decisions, \etc

Traditional system identification procedures rely on a mathematical, structured model that defines the relationship between system parameters and variables.
When considering dynamic physical systems, state-space models in the form of ordinary differential equations (ODEs) offer a mathematical formulation to establish this relationship~\cite{yassin_2013_recent}.
Such dynamic system models are often referred to as white-box models. White-box models help control engineers to make sure an industrial process is controlled optimally. However, the limited set of parameters in white-box models fail to fully align with real-world processes. 
 
Recently, data-driven machine learning (ML) based approaches have attracted attention for analyzing or understanding complex processes. ML based system identification has been used to build black-box models that define the relationship between control inputs and unknown parameters~\cite{chiuso_2019_mlSLreview}. Such black-box models, however, lack the physical understanding of the system and can result in outputs that are physically improbable when the models fail to generalize well. 
To overcome this, grey-box ML based methods have been used in recent years, where the process dynamics are integrated with the data-driven approach~\cite{Bohlin_2010_GBPI}.

Our objective in this paper is to present grey-box neural networks that are physics-informed, \ie which adhere to prior physical knowledge.
We particularly present a methodology for neural networks to follow a first order differential equation that describes the physical process.
As an exemplary use case, we focus on an induced draft cooling tower~\cite{VILJOEN_2018_coolingtower}.
Specifically, we devise and evaluate two state-of-the-art grey-box models, applied to the induced draft cooling tower: physics informed neural networks ({\PyNN}) and physics informed long-short term memory networks ({\PyLSTM}). 
These models enforce the system's dynamics onto the neural network, in contrast to a black-box neural network ({\NN}) model that follows a free structure. {\PyLSTM} is based on a recurrent neural network ({\RNN}) architecture. 
{\RNN}s are types of artificial neural networks that use the output of previous steps as inputs of the current step. Such architecture helps to learn the sequential dependencies and is beneficial for time-dependent modeling.  
Furthermore, we define a formulation to estimate operational flexibility based on system response from these identified models.

Our main contributions in this paper are, that we
\begin{enumerate*}[label=(\roman*)]
    \item devise the architecture for two novel data-driven grey-box modeling approaches, based on physics informed networks, \ie {\PyNN} and {\PyLSTM} (\sref{sec:data_driven_models}),
    \item extend the white-box model for an {\ECS} to calculate operational flexibility based on two proposed metrics (\sref{sec:cooling_tower}), and
    \item compare the accuracy and evaluation time of physics informed networks ({\PyNN} and {\PyLSTM}) with those of both white-box as well as black-box (\eg neural network {\NN}) models.
\end{enumerate*}
For the performance comparison, we simulate data from a white-box model (\sref{sec:ExperimentDesign}) and train the data-driven models based on designed experiments to evaluate the potential of data-driven grey-box models for system identification and address the following questions (\sref{sec:results}):
\begin{enumerate}[label=\textbf{(Q\arabic*)},ref={Q\arabic*},topsep=0pt,leftmargin=*,itemsep=0pt,labelsep=6pt]
    \item  \label{q:greyboximprovement} How do the grey-box models perform compared to the black-box and white-box models, in terms of accuracy?
    \item \label{q:LSTMimprovement} Does using a recurrent neural network architecture decrease the estimation error of the grey-box models ({\PyLSTM} compared to {\PyNN})?
    \item \label{q:Impacttraindata} What is the impact of amount of training data on
    \begin{enumerate*}[(i)]
    \item the estimation error,
    \item the optimization, and
    \item the training and evaluation time 
    \end{enumerate*}
    of the data-driven models?
    \item \label{q:opflex} What is the estimated operational flexibility in {\ECS} using data-driven models?
\end{enumerate}
\sref{sec:conclusions} summarizes conclusions and open issues that can be addressed in future work.
We note that the specific models and corresponding numerical performance results are for the selected representative use case of the evaporative cooling system, but the methodology followed to construct these models is applicable in general to various processes that can be approximated as a first order differential equation.
\section{Related work}
\label{sec:relatedwork}
\subsection{Machine learning in System Identification}
System identification in energy-intensive industrial processes such as an evaporative cooling system ({\ECS}) is a crucial step for control design, fault detection, flexibility exploitation, etc. Traditionally, white-box models are used for system identification.
For example, a mathematical model for simulation of refinery furnaces is proposed in~\cite{diaz2010mathematical}. The furnace model quantifies the main control variables for refinery services by combining a process model and a flue gas side model.
A model-based system identification for fault detection in a gas turbine is proposed in~\cite{Simani_2005_windturbinefault}, which identifies `residuals' that are the errors in the measured and estimated variables of the process.
As opposed to identification for fault detection or forecasting, we rather focus on identifying system output that can be exploited to build effective control for an energy-intensive industrial process (\ie by identifying system parameters that provide operational flexibility).

In contrast to~\cite{Simani_2005_windturbinefault}, which uses white-box models for system identification,
multiple data-driven black-box models are proposed in~\cite{Khayyan_2015_dynamicmodel} to predict the parameters of an energy-intensive thermal stabilization process in carbon fiber production.
A black-box long-short term memory ({\LSTM}) recurrent {\NN}~in~\cite{Wen_2019_LSTMpowerflactuations} provides accurate predictions of power fluctuations compared to real-time measurements.
In our work, we focus on data-driven \emph{grey-box} models, which are superior to the black-box models, as they include the process dynamics in the modeling approach and thus better adhere to physical constraints of the system (\ie by avoiding physically improbable outputs).

Grey-box machine learning models have been studied in the past. For example, \cite{raissi_2019_pinn} proposed a physics-informed neural network architecture that learns the system's output while respecting its physical laws (described using ODE/PDE). They evaluated these networks on systems described by the Schrodinger equation, the Navier–Stokes equation, \etc A grey-box model based on neural networks that respect Lagrangian mechanics is proposed in~\cite{Lutter_2019_DeepLN}.
Our work is different from these works in two aspects: 
\begin{enumerate*}[label=(\roman*),itemjoin={,\quad}]
    \item unlike~\cite{raissi_2019_pinn} and \cite{Lutter_2019_DeepLN}, we evaluate the grey-box models for a energy-intensive {\ECS}, and
    \item we propose {a newly designed} {\LSTM}\newtext{-}based recurrent neural network architecture for the grey-box modeling
\end{enumerate*}.
We evaluate our grey-box models for system identification in an induced draft cooling tower that is a part of an ECS.

\subsection{Evaporative Cooling System (ECS)}
An {induced draft} cooling tower is an essential part of the overall {\ECS} that provides opportunities to build effective control. A white-box model for the induced draft cooling tower based on first principles is proposed in~\cite{VILJOEN_2018_coolingtower}. A comparison of this white-box model and a black-box model based on an adaptive neuro-fuzzy interference system (\ie a neural network trained with an interference system based on fuzzy rules) in~\cite{GBeatens_2019_adaptive_neuro_fuzzy} shows that white-box models are superior for identifying basin temperature in the induced draft cooling tower. However, white-box models are computationally expensive and challenging to implement in a real-world setup. In this paper, we aim to overcome these challenges by defining grey-box physics informed networks ({\PyNN} and {\PyLSTM}) that
\begin{enumerate*}[label=(\roman*)]
    \item are computationally less  expensive than the white-box models defined in~\cite{VILJOEN_2018_coolingtower}, and
    \item respect the physical laws of the system, as opposed to the model defined in~\cite{GBeatens_2019_adaptive_neuro_fuzzy}
\end{enumerate*}.

Since flexibility is an abstract term, its quantification is a difficult task.
Approaches towards quantifying and exploiting operational flexibility follow simple methodologies. For example, in~\cite{aras2017application} technical flexibility constraints are estimated individually in energy-intensive processes, while ignoring their internal dependencies. Quantifiable features of demand-side flexibility (\eg time frame, storage capacity, interdependent jobs etc) are identified in \cite{barth2018comprehensive}. 
As opposed to using such a broad range of demand side parameters, we define operational flexibility using just two metrics, namely, state of charge and rate of charge, which represent the flexible characteristics of the system response (See further, \sref{sec:flexibility}).

\section{Induced Draft Cooling tower}
\label{sec:cooling_tower}
\begin{figure}[tb!]
    \centering
    \includegraphics[width=0.3\textwidth]{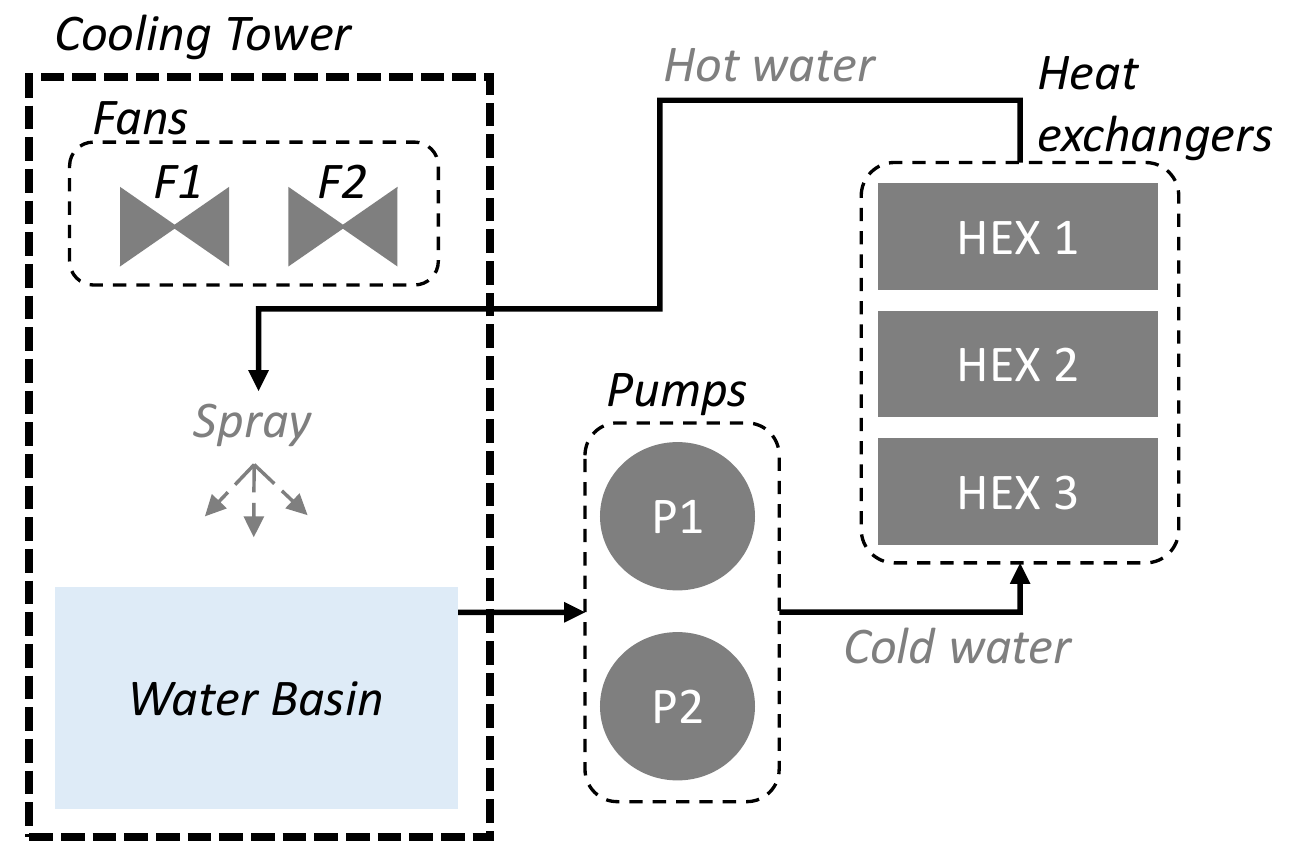}
    \caption{Simplified diagram of a evaporative cooling system}
    \label{fig:ECT_CoolingTower}
\end{figure}
{An induced draft cooling tower is a part of the {\ECS}}. A typical {\ECS} comprises fans and cooling water pumps, which are the primary energy consumers, as well as pipes, valves, and heat exchangers (HEXs).  \Fref{fig:ECT_CoolingTower} shows the simplified version of the considered {\ECS}, where the water is circulated and pumped using pumps to the set of HEXs.  They cool down the process in a closed-circuit. The cooling tower consists of fans, a hot water spray, and a water basin.

The physical state change from a fluid (water) to a gas (water vapor) requires latent heat. This heat is extracted from the water, resulting in a temperature decrease. This physical phenomenon is exploited within the cooling process by spraying warm water in a cooling tower, where fans induce a counter airflow to maximize the air-water interaction.  As a result, the water falling in the basin underneath has dropped in temperature and can be recirculated through the process. 
Together with the heat evacuation, also mass leaves the system in the form of water vapor. 

During this process, the basin temperature should remain within a specified range. It can be controlled using the control inputs (fan powers) and can be employed to calculate the process's operational flexibility.
We will estimate the operational flexibility by simulating and predicting the water basin temperature response to fan power changes. Different methods can be adopted to model the basin temperature based on the fans' power, including mathematical models and data-driven models.
In this section, a white-box model of the induced draft cooling tower based on previous research~\cite{GBeatens_2019_adaptive_neuro_fuzzy} is extended to estimate the operational flexibility of the process. 
\subsection{White-box model}
\label{sec:white_box_model}
A white-box thermodynamic model of the cooling tower system is developed based on the laws of heat transfer and mass conservation \cite{GBeatens_2019_adaptive_neuro_fuzzy}. This model is presented in \eref{eqn:ECT_tbbyt}, where {\BasinTemp} is the basin temperature. This model is defined using the process heat {\Qp} [W], the cooling capacity {\Qt} [W], the specific heat capacity of water {\cw} [kJ/kg·k], the water volume {\Vw} [m$^3$] and the water density {\rhow} [kg/m$^3$].
\begin{equation}\label{eqn:ECT_tbbyt}
    \frac{\partial T_{b}}{\partial t} = \frac{Q_p - Q_t}{c_w \cdot V_w \cdot \rho_w}
\end{equation}
The process heat {\Qp} is expressed as in \eref{eqn:ECT_Qp}, where {\mwdot} is the mass flow of water [Kg/s], {\ProcessTemp} is the process water temperature [K] and {\BasinTemp} is the water basin temperature [K]. Note that the {\mwdot}, \ie mass flow of process water, depends on the industrial process.
\begin{equation}\label{eqn:ECT_Qp}
    Q_p = \Dot{m}_w \cdot c_{w} \cdot ( T_p - T_b)
\end{equation}
The cooling capacity {\Qt} is expressed as in \eref{eqn:ECT_Qt}, where {\madot} is the mass flow of air [Kg/s], {\Henter} and {\Hleave} are the entering and leaving air enthalpy [kJ/kg·K], respectively.
\begin{equation}\label{eqn:ECT_Qt}
    Q_t = \Dot{m}_a \cdot ( H_{a,e} - H_{a,l})
\end{equation}
Enthalpies can be calculated using humidity rates and temperature \cite{Roland_2011_wetbulbtemp}. The entering air enthalpy depends on the ambient air temperature, and the leaving air enthalpy depends on the leaving air temperature. The mass flow rate of air depends on the power supplied to the fans and is expressed in \eref{eqn:ECT_ma}.
\begin{equation}\label{eqn:ECT_ma}
    \Dot{m}_a = \sqrt{ \frac{ 2 \cdot P_f \cdot \rho_a \cdot A^2_{\text{fr}} \cdot \eta_{\text{fan}} \cdot \eta_{\text{motor}} }{ 6.5 + K_{el} + 2 \cdot \frac{A^2_{\text{fr}}}{A^2_{\text{fan}}} }}
\end{equation}
where {\FanPower} is the electric fan power [W],  $A_{\text{fr}}$ is the tower frontal area [m$^2$], $A_{\text{fan}}$ is the fan area [m$^2$], $\eta_{\text{fan}}$ is the fan efficiency [\%], $\eta_{\text{motor}}$ is the motor efficiency [\%] and $K_{el}$ is the eliminator coefficient (if unknown, it is set to 1). The air density is represented by {\rhoa} and is calculated for the mixture of dry air and water vapor.
In a cooling tower, reliable operation is ensured if the basin temperature is within a range defined using the minimum and maximum operational requirements (\eref{eqn:Constraints_Tb}). Operational flexibility offered by the cooling tower based on the range of {\BasinTemp} is defined next.
\begin{equation}\label{eqn:Constraints_Tb}
  T_{b,\text{min}} \leq T_{b} \leq T_{b,\text{max}}
\end{equation}
\subsection{Operational Flexibility}
\label{sec:flexibility}
To quantify operational flexibility, we adopt the framework of~\citet{Ulbig_2015_op_flex}, who represent the system as a virtual battery. 
This virtual battery is defined by representing the process in the form of a power node~\cite{Heussen_2010_power_nodes}. 
The advantage of this approach is that it is independent of system configurations (\eg  the number of fans), and can generalize to any system represented as a virtual battery.

To quantify the operational flexibility in an evaporative cooling tower, we represent it as a thermal battery, where power is fed into the process through the fans that feed air into the tower. The operational flexibility of this thermal battery is given by two metrics, namely, 
\begin{enumerate*}[label=(\roman*)]
    \item State of Charge ({$\SoC$}), and, 
    \item Rate of Charge ({$\RoC$}).
\end{enumerate*} 
 The {$\SoC$} represents the flexibility characterized by the state of operation, and is calculated using the basin temperature {\BasinTemp}. The {$\SoC$} is defined in \eref{eqn:ECT_SoC} where $1/\eta$ is the efficiency. The range of {$\SoC$} is from 0 to $(T_{b,\text{max}}-T_{b,\text{min}})/{\eta}$. The {$\RoC$} represents the flexibility offered charging/discharging, \ie the transition speed between states of operation. {$\RoC$} is defined in \eref{eqn:ECT_RoC} 
\begin{equation}\label{eqn:ECT_SoC}
    \SoC = {(T_{b}-T_{b,\text{min}})}/\eta
\end{equation}
\begin{equation}\label{eqn:ECT_RoC}
    \RoC = \frac{1}\eta \frac{ \partial T_{b}}{ \partial t}
\end{equation}
Compared to~\cite{Ulbig_2015_op_flex}, we thus define new operational flexibility metrics, \ie {$\SoC$} and {$\RoC$}, that are used specifically for energy intensive industrial processes such as the exemplary induced draft cooling tower. 
For the cooling tower, {$\SoC$} provides the nominal cooling production capacity and {$\RoC$} provides the cooling capacity variation relative to the power consumption.


\section{Data-Driven Models}
\label{sec:data_driven_models}
Data-driven models offer an alternative to white-box dynamic models (\sref{sec:white_box_model}) to construct behavioral input-output models.  
Data-driven models are faster to evaluate compared to white-box models. 
A data-driven model is also efficient in terms of {space complexity} and can be saved and evaluated in real-time. 
These models can be trained on either real-world data or simulated data. 
This section presents typical architectures for the black-box (\sref{sec:blackboxmodels}) and grey-box (\sref{sec:greyboxmodels}) data-driven models that are employed to approximate the time-varying relationship between the control inputs and system responses, specifically for the considered {\ECS}.
\sref{sec:LossFun} summarizes the loss functions used to train the models.
Finally, \sref{sec:OpFlexestimated} presents how the operational flexibility is estimated using the data-driven models.

\subsection{Black-box models}
\label{sec:blackboxmodels}
\begin{figure}[!tb]
    \centering
    \includegraphics[width=0.38\textwidth]{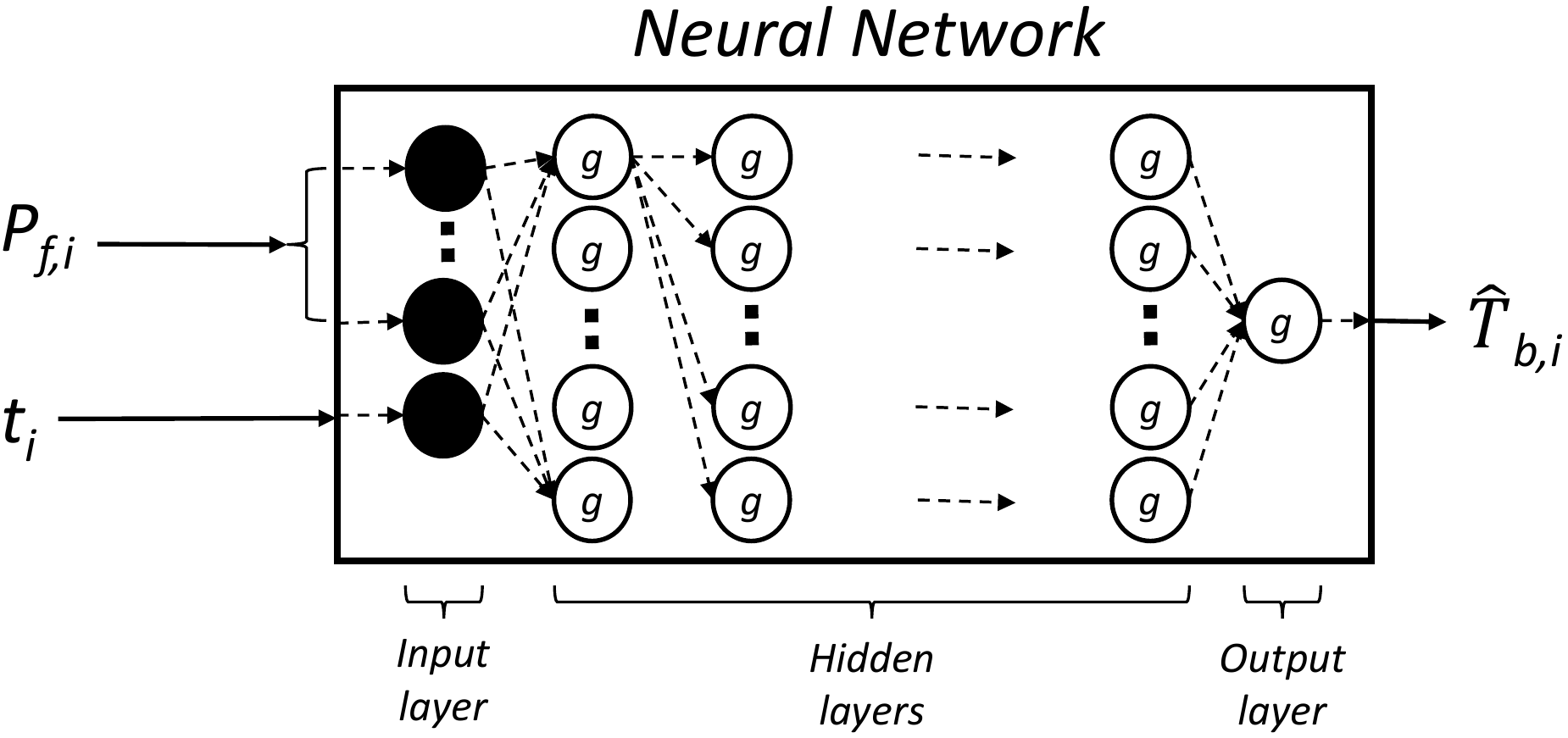}
    \caption{\small{Black-box {\NN} (Not all connections between cells are shown to keep the figure legible)}}
    \label{fig:ECT_NN}
\end{figure}
\begin{figure}[!tb]
    \centering
    \includegraphics[width=0.38\textwidth]{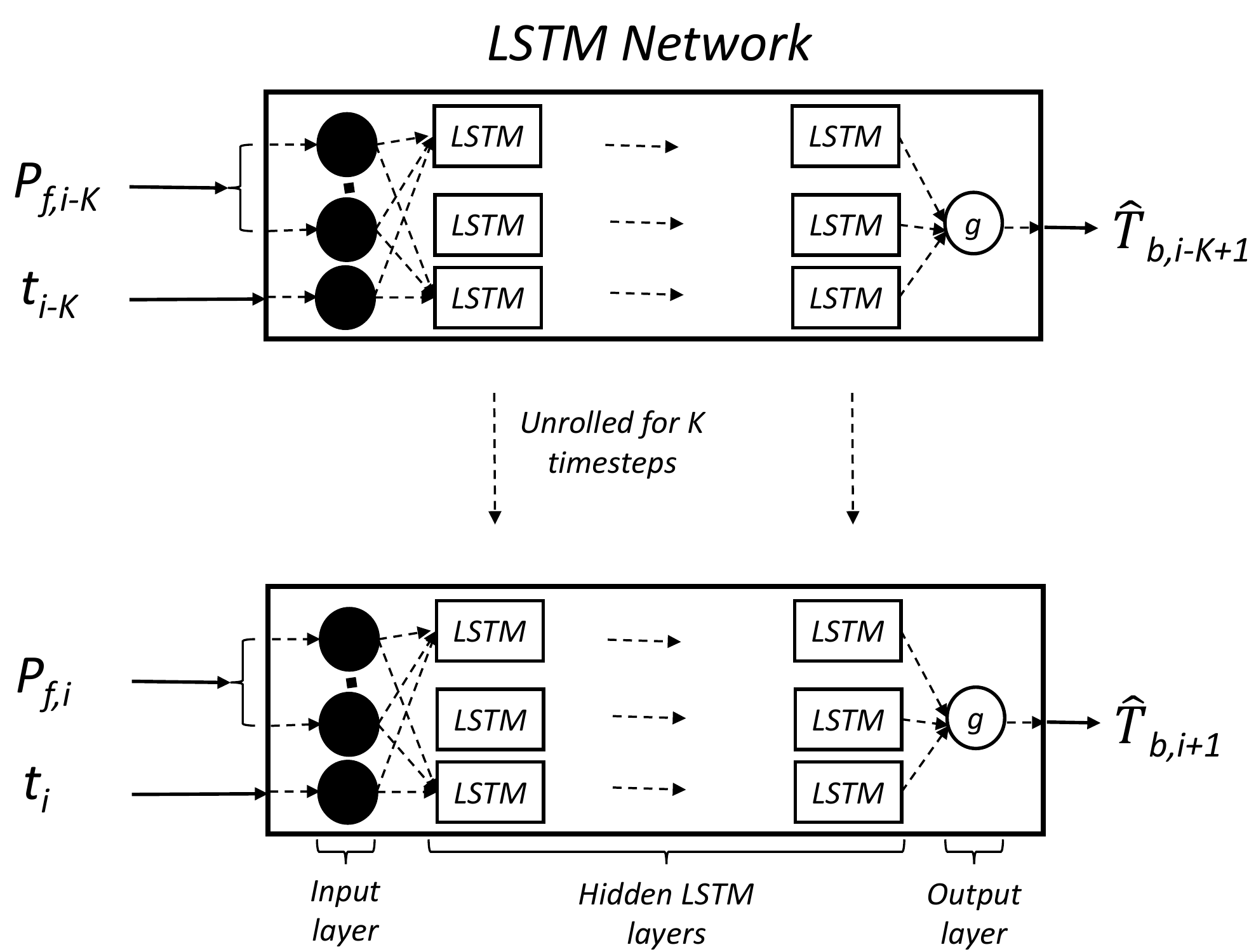}
    \caption{\small{Black-box {\LSTM} unrolled in time for K timesteps (Not all connections between cells are shown to keep the figure legible)}}
    \label{fig:ECT_LSTM}
\end{figure}
To approximate the evaporative cooling system, 
we consider two black-box models:
\begin{enumerate*}[label=(\roman*),itemjoin={,\quad}]
    \item a typical Neural Network ({\NN}), and
    \item a typical Long short-term memory network ({\LSTM}).
\end{enumerate*}
Both black-box models designed with the control input being the fan power {\FanPower} and system response basin temperatures {\BasinTemp}. 
The basin temperature depends on the time of day (since ambient air temperature and its pressure change during the day). Thus we include the timeslot (\textit{t}) as model inputs, providing a temporal component to the model. The approximated basin temperature $\widehat{T}^\textit{NN}_{b}$ is obtained from a neural network (\textit{\NN}) with weights $\theta$, as stated by~\eref{eqn:NN_Tbasin}. The timeslot $t_i$ represents the time of day. For example, if the timeslot duration ({\tslot}) is 15 minutes, a 24 hour day will have 96 timeslots, \ie \textit{t$_i$} $\in \{1,2,\dots,96\}$. 
\begin{align}
  \widehat{T}^\textit{NN}_{b,i}= \textit{NN}({P_{f,i} , t_i ; \theta})
  \label{eqn:NN_Tbasin}
\end{align}
\Fref{fig:ECT_NN} provides the architecture of the mentioned {\NN} model, where \textit{i} represents the $i^{th}$ observation. The activation function for each cell is represented by \textit{g} and can be chosen depending on the problem (\eg sigmoid, softmax, ReLu, etc.)

An {\LSTM} network is a type of recurrent neural network where each layer has {\LSTM} cells characterized by a state that is updated based on the previous timestep state and current input, controlled through so-called forget gate, input gate and output gate~\cite{SHochreiter_1997_LSTM}. This network is used for temporal or sequential modeling, where the outputs of a step depend on the output of previous steps. The approximated basin temperature $\widehat{T}^\textit{LSTM}_{b}$ at timestep $t_i$ thus depends on the control inputs of previous $K$ timesteps (\eref{eqn:LSTM_Tbasin}).
To estimate the basin temperate at the $i^{th}$ timestep, we sequentially feed the control inputs from timestep $i-K$ to timestep $i$. The sequence length (\textit{K}) depends on the problem and can be chosen based on the past dependencies of the output.
\Fref{fig:ECT_LSTM} provides the architecture of the mentioned {\LSTM} model where we feed data for  $K$  timesteps and get the final output basin temperature for the $i$+1$^\textrm{th}$ timestep.

\begin{align}
  \widehat{T}^\textit{LSTM}_{b,i} =  \textit{LSTM}({P_{f,i}, t_{i} ,\dots, P_{f,i-K}, t_{i-K} ; \theta})
  \label{eqn:LSTM_Tbasin}
\end{align}

\subsection{Grey-box models}
\label{sec:greyboxmodels}
A grey-box model combines data-driven with physics-based models to improve the reliability of the model. 
Physics informed networks are used as grey-box models to approximate the behavior of a dynamic system. 
An induced draft cooling tower is characterized by~\eref{eqn:ECT_tbbyt}, where {$T_{b}$} is the system response which is approximated using a neural network.
\citet{raissi_2019_pinn} suggest to define a function $f$ by rewriting~\eref{eqn:ECT_tbbyt} as in~\eref{eqn:ECT_f}, thus allowing us to evaluate the adherence of the grey-box model to the system dynamics by utilizing $\widehat{f}$, as the approximation of ${f}$. 
The second term of this is defined as operator $H$ in~\eref{eqn:ECT_H},  
characterized by parameters $\beta$, which represents the parameters used in the calculation of $Q_p$  and $Q_t$. In our case of a cooling tower, $\beta$ includes constants such as specific enthalpy, density, the universal gas constant, \etc
We can calculate $Q_p$, using \eref{eqn:ECT_Qp}, based on {\BasinTemp} and $\beta$.
Similarly, $Q_t$ can be calculated with \eref{eqn:ECT_Qt} and \eref{eqn:ECT_ma} using {\BasinTemp}, $\beta$ and {\Weathertuple} (a tuple constructed from temperature, pressure, and humidity of ambient air). 

\begin{equation}\label{eqn:ECT_f}
    f \triangleq \frac{\partial T_{b}}{\partial t} - \frac{Q_p - Q_t}{c_w \cdot V_w \cdot \rho_w}
\end{equation}
\begin{equation}\label{eqn:ECT_H}
    H(T_b, W ; \beta) = \frac{Q_p - Q_t}{c_w \cdot V_w \cdot \rho_w}
\end{equation}
\subsubsection{Physics Informed Neural Networks ({\PyNN})}
\label{sec:PyNN_Py_based}
\begin{figure}
    \centering
    \includegraphics[width=0.38\textwidth]{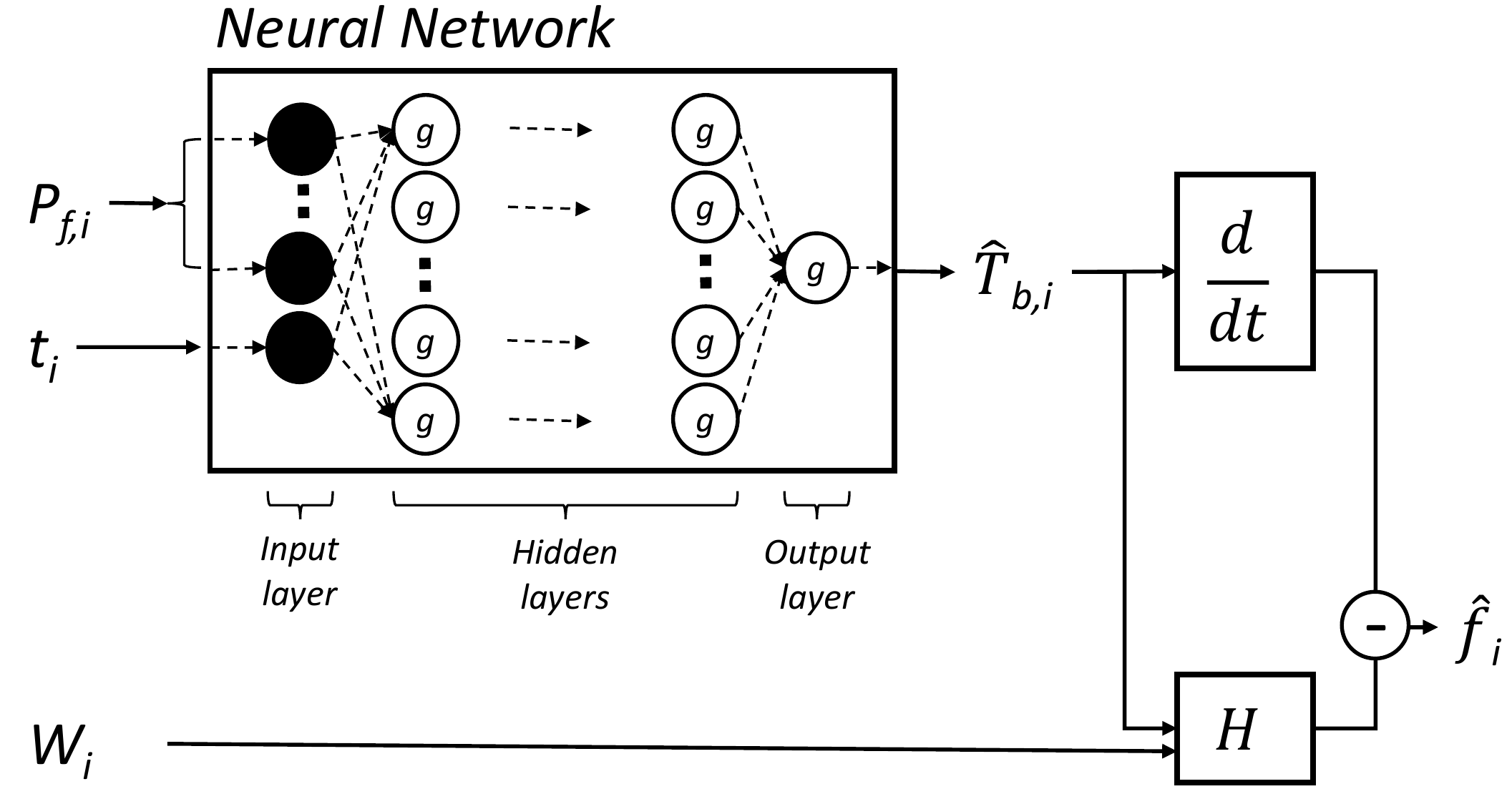}
    \caption{\small{Grey-box {\PyNN} (Not all connections between cells are shown to keep the figure legible).}}
    \label{fig:ECT_PyNN}
\end{figure}
Dynamics of the cooling tower can be enforced on the neural network architecture using \eref{eqn:ECT_f}. In \fref{fig:ECT_PyNN} we illustrate the {\PyNN} architecture, where \textit{i} is the $i^{th}$ observation. A neural network is used to approximate the basin temperature 
(\eref{eqn:PyNN_Tbasin}). The time derivative of the basin temperature is determined using auto differentiation. The function $\widehat{f}^\textit{PhyNN}$ is estimated using $\widehat{T}^\textit{PhyNN}_{b}$ and the weather data (W: Temperature, pressure and humidity) for each timestep,~\eref{eqn:PyNN_f}.

\begin{align}
  \widehat{T}^\textit{PhyNN}_{b,i}= \textit{PhyNN}({P_{f,i} , \textit{t}_i ; \theta})
  \label{eqn:PyNN_Tbasin}
\end{align}
\begin{equation}\label{eqn:PyNN_f}
    \widehat{f}^\textit{PhyNN}_{i} = \left(\frac{\partial \widehat{T}^\textit{PhyNN}_{b}}{\partial t}\right)_{i} - H(\widehat{T}^\textit{PhyNN}_{b,i}, W_i ; \beta)
\end{equation}
\subsubsection{Physics Informed {\LSTM} Networks (\PyLSTM)}
\label{sec:PyLSTM}
\begin{figure}[!tb]
    \centering
    \includegraphics[width=0.38\textwidth]{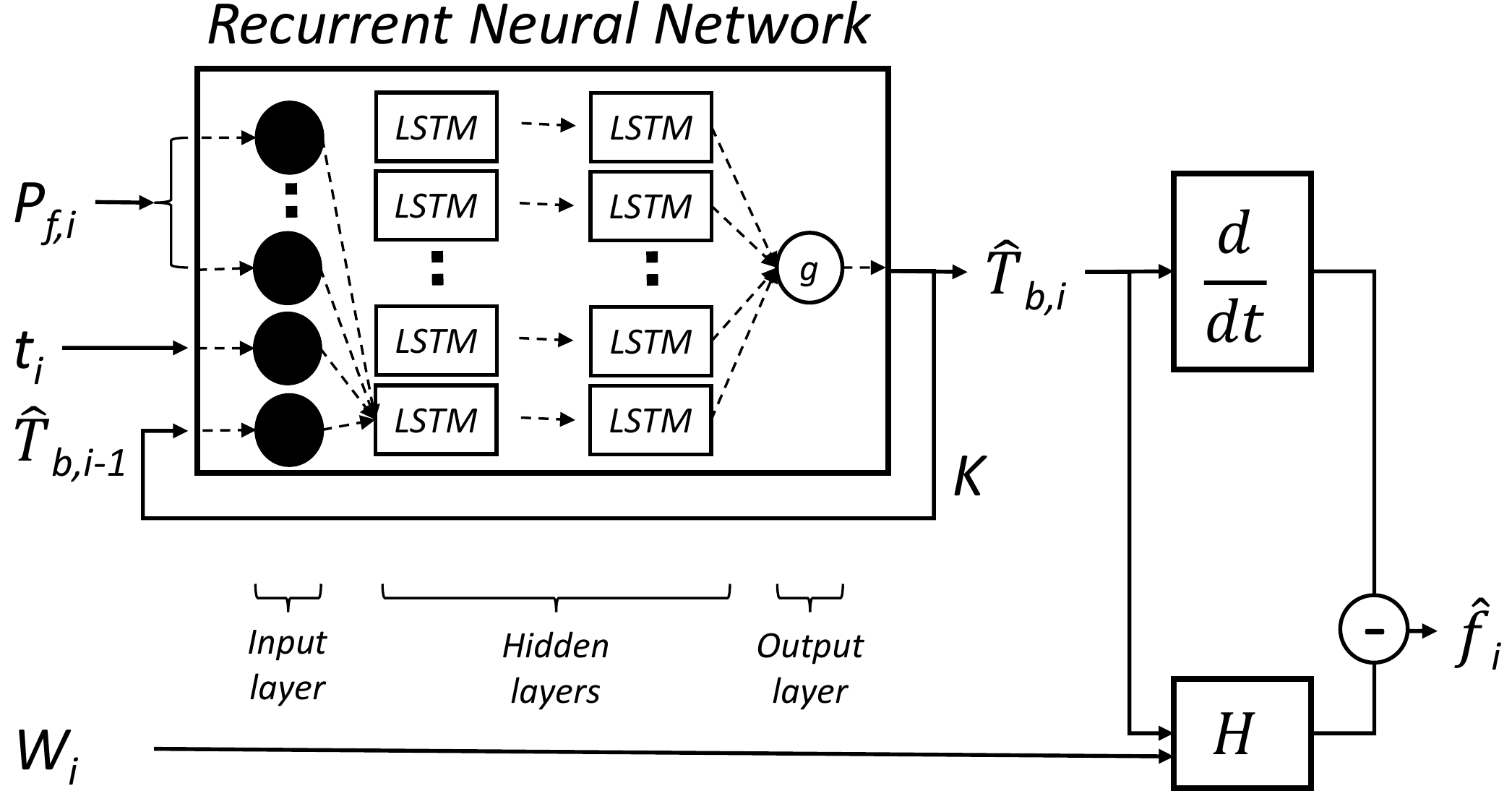}
    \caption{\small{Grey-box {$\text{PhyLSTM}_\textit{WF}$}. The physics informed {\LSTM} network with feedback from output to inputs of the {\LSTM} network.  (Not all connections between cells are shown to keep the figure legible).}}
    \label{fig:ECT_PyLSTM}
\end{figure}
The {\PyLSTM} model is designed to enforce the physics of the cooling tower on a typical {\LSTM} network at each timestep $i$.
The approximated value of basin temperature parameter $f$ at step $i$ is calculated using~\eref{eqn:PyLSTM_Tbasin}~and~\eref{eqn:PyLSTM_f}.

\begin{equation}
  \widehat{T}^\textit{PhyLSTM}_{b,i}= \textit{PhyLSTM}({P_{f,i}, t_i, \dots, P_{f,i-K}, t_{i-K};\theta})
  \label{eqn:PyLSTM_Tbasin}
\end{equation}
\begin{equation}\label{eqn:PyLSTM_f}
    \widehat{f}^\textit{PhyLSTM}_{i} = \left(\frac{\partial \widehat{T}^\textit{PhyLSTM}_{b}}{\partial t}\right)_{i} - H(\widehat{T}^\textit{PhyLSTM}_{b,i}, W_i ; \beta)
\end{equation}
Similar to the {\PyNN}, the time derivative is calculated using auto differentiation and $H$ is calculated using the weather data (${W_{i}}$) and the output of the network ($\widehat{T}^\textit{PhyLSTM}_{b,i}$).

LSTM is a widely known RNN architecture for time series modeling which updates weights by error propagation through time using LSTM cell states. The LSTM cell architecture consists of connections (known as gates) that manipulate inputs, \eg $P_{f}$, to estimate the output, \eg $T_{b}$, and update the cell state in each 
recurrent step~\cite{SHochreiter_1997_LSTM}.
This updated cell state feedback is further used in the next recurrent timestep. 
NARX models offer an alternative way of time series modeling, through using direct feedback of the output.
As opposed to a linear chain of LSTM cells, a NARX model uses a simple feed-forward network to estimate outputs~\cite{NARXNeuralNetworks}.
A NARX uses direct feedback connections: the output of the past timestep is fed as input of the current timestep, but without having other state feedback (\cf the hidden state of an LSTM cell).
The output feedback in these models has proven its effectiveness in  accurate long-term time series modeling~\cite{Jose_2008_NARXeval}.
In PhyLSTM, physics is enforced on the output $T_b$ estimated by the LSTM network, where LSTM cell states are propagated through time. We note that the basin temperature ($T_b$) of the current timestep depends on its past values. Thus, to fully explore the potential of PhyLSTMs, in addition to cell state feedback, we introduce output feedback for time series modeling of $T_b$.

We design two {\PyLSTM} architectures based on the feedback from the output of the network to its input. More specifically,
\begin{enumerate*}[label=(\roman*)]
    \item an architecture \emph{with} feedback {$\text{PhyLSTM}_\textit{WF}$}, where to estimate $\widehat{T}^\textit{PhyLSTM}_{b}$ at $i^{th}$ timestep, the estimated basin temperature of $(i-1)^{th}$ timestep is fed as input to the {\LSTM} network, and
    \item an architecture \emph{without} feedback {$\text{PhyLSTM}_\textit{WOF}$}.
\end{enumerate*}
\Fref{fig:ECT_PyLSTM} describes the architecture of {$\text{PhyLSTM}_\textit{WF}$}, with the feedback of previous basin temperature for a single timestep. During training, we unroll the network for $K$ timesteps similar to the black-box {\LSTM} network~\fref{fig:ECT_LSTM}. In case of {$\text{PhyLSTM}_\textit{WOF}$}, the output feedback connection in the {\LSTM} network is removed. 
\subsection{Training and Loss Functions}
\label{sec:LossFun}
For the black-box {\NN} and {\LSTM} models, the weights ($\theta$) of the networks are optimized by minimizing the loss function defined using mean squared error ({\MSE}) in estimating the basin temperature {\BasinTemp}.
For example, \eref{eqn:NN_loss} provides this loss function for {\NN}, where the approximated value of basin temperature $T_{b,i}$ is represented by $\widehat{T}^{N\!N}_{b,i}$. The total number of observations in a batch is represented by \textit{B}.
The second term in \eref{eqn:NN_loss} is $\mathcal{L}2$ regularization, which prevents the network's weights from increasing excessively. The hyper-parameter $\lambda$ controls the importance of the regularization term.

\begin{align}
  \mathcal{L}^\textit{NN}(\theta)&= \frac{1}{B} \sum\limits_{i=1}^{B} (T_{b,i} - \hat{T}^\textit{NN}_{b,i})^2 + \frac{\lambda}{B} \sum\limits_{i=1}^{B} |\theta_i|^2 
  \label{eqn:NN_loss}
\end{align}
\begin{align}
  \mathcal{L}^\textit{PhyNN}(\theta) = &\frac{1}{B} \sum\limits_{i=1}^{B} \Biggl\{ \left(T_{b,i} - \widehat{T}^\textit{PhyNN}_{b,i}\right)^2 + \left(\widehat{f}^\textit{PhyNN}_{i}\right)^2 \Biggr\}\notag \\
  &+ \frac{\lambda}{B} \sum\limits_{i=1}^{B} |\theta_i|^2 
  \label{eqn:PyNN_loss}
\end{align}

Loss functions used to train the neural networks in grey-box models are based on the error in estimating {\BasinTemp}: ($T_{b,i}~-~\widehat{T}^\textit{PhyNN}_{b,i}$), and the error in enforcing system dynamics, \ie {\f}: ($f_{i}~-~\widehat{f}^\textit{PhyNN}_{i}$).
We train the networks by minimizing the total mean squared error ({\MSE}), \ie summing the MSEs of {\BasinTemp} and {\f} \cite{raissi_2019_pinn}.
Adding the {\MSE} of {\f} ensures that the neural network maximally adheres to the dynamic system's physics.
In~\eref{eqn:PyNN_loss}, we give the loss function for {\PyNN} based on weights $\theta$. 
Estimated values of {\BasinTemp} and {\f} for the $i^{th}$ observation are represented by $\widehat{T}^\textit{PhyNN}_{b,i}$ and $\widehat{f}^{P\!h\!y\!N\!N}_{i}$ respectively. The last term is $\mathcal{L}2$ regularization. The equation is simplified based on the true value of $f_i$ = 0 (\eref{eqn:ECT_f}).

While training the {\PyLSTM}, 
we pass $K$ sequential inputs through the network. The full set of observations in a batch ($B$) are divided into $C$ parts, where each part comprises $K$ sequential inputs. The total number of such parts is given by $C$ = $\floor{B/K}$. Thus, the total number of observations used to train {\PyLSTM} is \textit{CK}.
\eref{eqn:PyLSTM_loss} provides the loss function for the {\PyLSTM} model, where mean square errors are calculated for all parts and all timesteps.
$\mathcal{L}2$ regularization is specifically useful to avoid overfitting an {\LSTM} network trained using `teacher-forcing'.\footnote{During training true value of the previous timestep is used rather than the estimated value. Refer to \sref{sec:Training} for further details.}

\begin{align}
\quad
  \mathcal{L}&^\textit{PhyLSTM}(\theta) = \notag \\
  &\frac{1}{CK} \sum\limits_{c=1}^{C}\sum\limits_{k=1}^{K} \Biggl\{ \left(T_{b,k} - \widehat{T}^\textit{PhyLSTM}_{b,k}\right)^2 
  +\left(\widehat{f}^\textit{PhyLSTM}_{k}\right)^2 \Biggr\}_{c} \notag \\
  &+ \frac{\lambda}{CK} \sum\limits_{i=1}^{CK} |\theta_i|^2
  \label{eqn:PyLSTM_loss}
\end{align}
\begin{algorithm}[t]
\small
 \caption{Training \PyNN~and~\PyLSTM}
 \label{Algo:PyNetworkTraining}
 \begin{algorithmic}[1]
 \renewcommand{\algorithmicrequire}{\textbf{Input:}}
 \renewcommand{\algorithmicensure}{\textbf{Output:}}
 \REQUIRE $\mathcal{D}$ = $\{(P_{f}, t, W)_i$,$~(T_{b})_i\}_{i=1}^{B}$
 \ENSURE $\mathcal{M}$ (Trained model)
 \STATE $\mathcal{M}$ = Model $\in \{\textit{\PyNN}, \textit{\PyLSTM\}}$;
 \STATE $\theta$ = weights of the network in $\mathcal{M}$;
 \WHILE{not stopping condition}
     \FOR {all i}
        \STATE $\widehat{T_b}$ = $\mathcal{M}$($P_f, t ;  \theta$);\\
        \STATE $\widehat{f}$ = $\mathcal{M}$($P_f, t, W ;  \theta$);
        \STATE $\theta \leftarrow \theta - \alpha \frac{d L^{\mathcal{M}}(\theta)}{d\theta}$; 
        \footnotesize{\color{gray}{\COMMENT{Update weights using $\widehat{T_b}$ and $\widehat{f}$}}}
     \ENDFOR
 \ENDWHILE 
 \RETURN $\mathcal{M}$ 
 \end{algorithmic} 
 \end{algorithm}

Pseudocode for training the physics informed networks is provided in \algoref{Algo:PyNetworkTraining}, where we represent a physics informed network by $\mathcal{M} \in \{\PyNN,\PyLSTM\}$.
The loss $  \mathcal{L}^{\mathcal{M}}(\theta)$ is given by~\eref{eqn:PyNN_loss} or~\eref{eqn:PyLSTM_loss}, based on the chosen model $\mathcal{M}$.
\subsection{Operational Flexibility Identification}
\label{sec:OpFlexestimated}
Using the physics informed networks, we can get the estimated values of the basin temperature {$\widehat{T}_{b}$} and its derivative $\frac{\widehat{dT_{b}}}{dt}$ (\sref{sec:greyboxmodels}).
These are used to estimate operational flexibility metrics.
For a timeslot $i$, we can estimate the cooling capacity $\widehat{\SoC}_i$ using \eref{eqn:pihat}, and cooling variations capacity $\widehat{\RoC}_i$ using \eref{eqn:rhohat}.
This quantitatively estimated flexibility can be exploited to design control, make cost decisions \etc

\begin{equation}
            \widehat{\SoC}_{i} = {\left(\widehat{T}_{b,i}-T_{b,\text{min}}\right)}{/\eta}
            \label{eqn:pihat}
\end{equation}
\begin{equation}
            \widehat{\RoC}_i = \frac{1}{\eta} \left( \frac{\widehat{dT_{b}}}{dt} \right)_i 
            \label{eqn:rhohat}
\end{equation}
\section{Experiment Design}
\label{sec:ExperimentDesign}
\Fref{fig:ECT_CoolingTower} provides a simplified version of the evaporative cooling system, which consists of a simple cooling tower.
A typical real-world evaporative cooling system consists of multiple fans, pumps, and heat exchangers, where sensors are employed to collect data about different parameters of the process.
However, such data often {suffers} from missing or noisy observations. Training data-driven models {requires} a reliable source of data, which is not readily available for research purposes. To overcome this, we simulate the data using the white-box based thermodynamic model (\sref{sec:white_box_model}), and  use this simulated data to train the data-driven models.

Data is generated for a cooling tower with two fans (F1 and F2) and one water basin. Controlling these fans changes both the electricity consumption of the cooling system and the water basin temperature. In this section, we describe data simulated using the dynamic modeling of the induced draft cooling tower. 
A detailed description of the training and evaluation of these data-driven models is also covered in this section.
\subsection{Data}
\label{sec:data}
Using the white-box model defined in \sref{sec:white_box_model} we simulate the data for each minute from Jan $1^{st}$ 2017 to $31^{st}$ Dec 2017, \ie 60 minutes $\times$ 24 hours $\times$ 365 days = 525,600 observations. Thus, the duration of each timeslot {\tslot} = 1 minute and we create 1,440 timeslots {\timeslot} for each day of 24 hours. Each day starts from midnight (00:00) and ends 24 hours later (23:59) and we represent each timeslot as a decimal value \timeslot~$\in$~$[0.00, 24.00)$ (for example, {\timeslot} = 1.5 means 01:30 AM). {We choose a 1 minute timeslot because, 
\begin{enumerate*}[label=(\roman*)]
    \item it provides sufficient training data compared to larger timeslots (\eg 15\,min)
    \item time-varying dependencies of the dynamic system are better captured compared to smaller timeslots (\eg 1\,s)  
    \item \textcolor{black}{control for such systems operates on minute basis, and our models can assist in developing such control.}
\end{enumerate*}
}

Power supplied to both fans ($P_{F1}$ and $P_{F2}$)  is chosen randomly from $\{100,150,200,250\}$ after a random duration. Gaussian noise is added to the simulated {\BasinTemp} to get realistic training data for neural networks. This noisy basin temperature is represented by {\BasinTempNoise}. 
Weather data (ambient temperature, pressure, and humidity) is collected for 2017 and used for calculating {\Qt} \cite{weather2017}. We observe that weather patterns have a daily seasonality, given by lower ambient temperatures during the night and higher ambient temperatures during the day. The timeslot is included in the data to enable modeling these time-dependent characteristics of the process. 

\begin{figure}[!tb]
    \centering
    \includegraphics[width=0.49\textwidth]{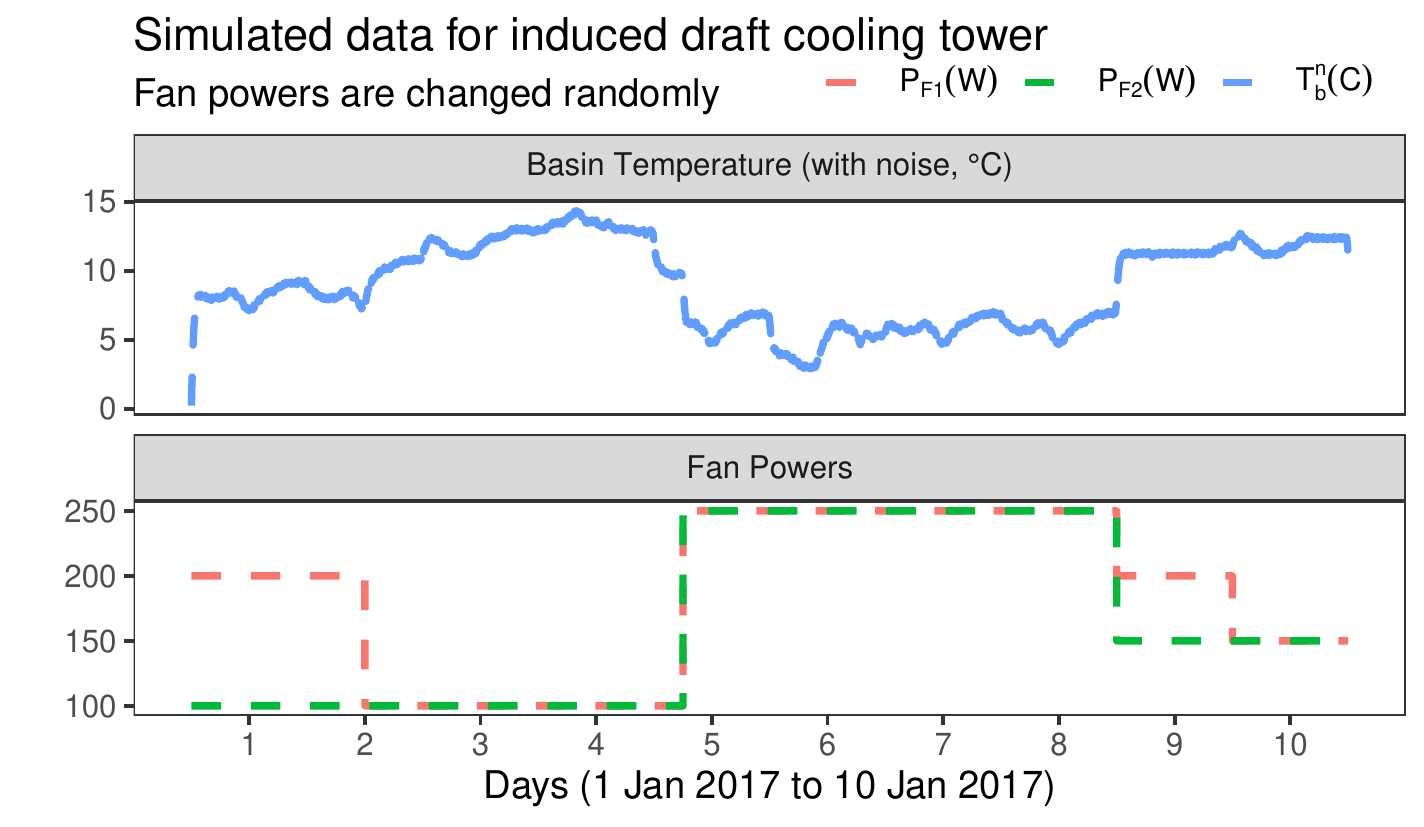}
    \caption{Simulated data for the first 10 days of 2017.}
    \label{fig:Simulated_data_10days}
\end{figure}

The process heat depends on the process water temperature that comes from the heat exchangers and is typically dependent on the industrial process's specification. 
We can safely assume a constant value for process heat {\Qp} to be 33\,W.
Temperature range is assumed to be 9$^\circ$K.
The python package `simpy' is used for simulation. 
\tref{table:Simulated_data_table} shows a small sample of the simulated data where time is represented by $t_{n}$ along with fan powers ($P_{F1}$ and $P_{F1}$) and basin temperature ($T^{n}_{b}$).
\Fref{fig:Simulated_data_10days} shows the first 10 days (Jan 1$^\textrm{st}$ to 10$^\textrm{th}$ 2017) of the data that was simulated using the white-box model. All data-driven models are evaluated using this simulated data.

The operational limits of {\BasinTemp} are assumed to be from $T_{b,min}$~= 9$^\circ$C to $T_{b,max}$ = 36$^\circ$C, and the average efficiency is set to $1/\eta$ = 0.6~\cite{GBeatens_2019_adaptive_neuro_fuzzy}. Thus, the range of the operational flexibility metric $\SoC$ given by \eref{eqn:ECT_SoC} is  [0, 16.2]. The $\RoC$ value given by \eref{eqn:ECT_RoC} depends on the characteristics of the dynamic system \eg number of fans, capacity of basin, \etc For experimental purposes, the range of $\rho$ is assumed to be [$-$10,~10].

\begin{table}[!tb]
    \begin{tabular*}{.45\textwidth}{@{\extracolsep{\fill}}c|c|c|c|c|c}
      {$t_{n}$} & {$P_{F1}$} & {$P_{F2}$} & {\Qp} & {\BasinTemp} & {\BasinTempNoise}\\
      {(time)} & {(W)} & {(W)} & {(W)} & {(K)} & {(K)}\\
      \hline
      0.000694 & 100 & 100 & 33 & 10 & 10.020\\
      0.001388 & 100 & 100 & 33 & 10 & 10.023\\
      \ldots & \ldots & \ldots & \ldots & \ldots & \ldots
    \end{tabular*}
    \caption{Simulated data ({\BasinTempNoise} is noisy basin temperature).}
    \label{table:Simulated_data_table}
\end{table}
\normalsize

\subsection{Training}
\label{sec:Training}
Two black-box models ({\NN} and {\LSTM}) and three grey-box models ({\PyNN}, {\PyLSTM$_\textit{WF}$} and {\PyLSTM}$_\textit{WOF}$) are trained and evaluated. The noisy temperature of basin ({\BasinTempNoise}) is used as an output to train these models, and the power of the fans ($P_{F1}$ and $P_{F1}$) and timeslot ({\timeslot}) are used as inputs.
Time ({\timeslot}) is fed as a decimal value~({\timeslot}$~\in~$(0.00, 24.00), as described in \sref{sec:data}).

\begin{figure}[!tb]
    \centering
    \includegraphics[width=0.4\textwidth]{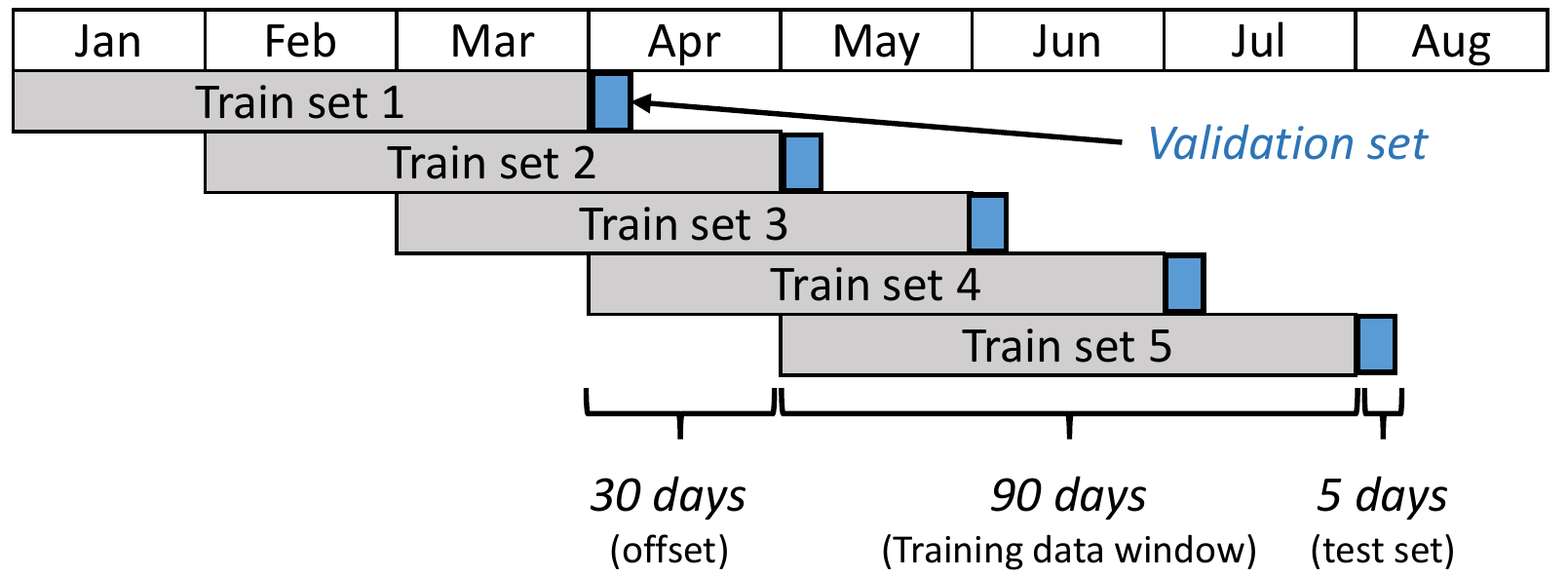}
    \caption{Walk forward-moving window validation for training period of three months (90 days). }
    \label{fig:trainvalsets}
\end{figure}

We use the simulated data for 2017 described in \sref{sec:data} to train the data-driven models. We define multiple training and validation sets using a walk forward moving window validation (\fref{fig:trainvalsets}). This validation is conducted because the basin temperature is time series data. All models are trained for \textit{training period} $\in \{1, 3, 5, 7\}$ months of training data (with 30 days in each month) with a moving window of 30 days, and a total of five validation sets are applied for each number of training months. We train models on different numbers of training months to evaluate the impact of training data on the model performance.
To train the model, we select data in the training duration from the 2017 data, and use~\algoref{Algo:PyNetworkTraining} to optimize the weights of the model.
Each trained model is evaluated for the next five days immediately following the end of the training period. 

All models are defined using TensorFlow in Python. The weights of the models are optimized by minimizing the loss functions defined in~\sref{sec:LossFun} using ADAM optimization for 1,000 iterations and a learning rate of 0.001. We use the Limited-memory BFGS optimization algorithm provided by the  Python library Scipy with a maximum of 5,000 function evaluations to complete the training. The remaining parameters of the optimizers' are set to their default values.

Black-box and grey-box models have fully connected networks. Each network has three input nodes, one output node, and two hidden layers with 16 neurons in each hidden layer.
In the case of {\PyLSTM$_\textit{WF}$}, one extra input node is added to feed the output of the previous timestep.
For all LSTM-based models (black-box {\LSTM} as well as grey-box {\PyLSTM$_\textit{WOF}$}) and {\PyLSTM$_\textit{WF}$}), 
the hidden layers have 16 {\LSTM} cells each. Sigmoid activation is used for all networks. 
We choose sigmoid because, we observed that ReLU activation made the {\LSTM} model diverge~\cite{ReLU_LSTM_Paper}. Model performances were compared for tanh and sigmoid activation functions {with one, two and three hidden layers}, where we note that sigmoid {with two hidden layers} provided faster and more accurate convergence.

Physics informed {\LSTM} networks are optimized for the recurrence of \textit{k} = 60 timesteps (60\,min). The RNN is trained using the `teacher-forcing' algorithm introduced in~\cite{Williams_1989_TeacherforcingRNN} where we use the previous timesteps' ground truth to train the network, \ie during training, the real value of the previous timestep is used as input as opposed to the value estimated from {\PyLSTM}. While making predictions from {\PyLSTM}, we feed the data for the last $k$ timesteps to make the prediction. The time derivative of the basin temperature is calculated using the auto differentiation functionality provided by the TensorFlow platform. 

\subsection{Evaluation}
\label{sec:Evaluation}
We evaluate the performance of the data-driven models using absolute errors encountered in estimating the system response, \ie the basin temperature. The absolute error for each timestep for a data-driven model $m$ is defined in \eref{eqn:ECT_abs_error}, where $T_{b,i}$ is the basin temperature simulated using the white-box model, and $\widehat{T}^{m}_{b,i}$ is the estimated basin temperature (using data-driven model $m$~$\in \{\text{NN},~\text{LSTM},~\text{PhyNN},~\text{PhyLSTM}_\textit{WOF},~\text{PhyLSTM}_\textit{WF}\}$).

\begin{equation}
	\label{eqn:ECT_abs_error}
	|e_i|^{m}  = |T_{b,i} - \widehat{T}^{m}_{b,i}| 
\end{equation}

Absolute errors are calculated for all models trained on different \textit{training periods} ($\in \{$1, 3, 5, 7$\}$ months) for each timestep in the immediate next 5 days, \ie for the next 7,200 minutes (timesteps). We also compare the training time for all models, \ie the time it takes to run each iteration of optimization.
The performance of data-driven models is compared for three periods of prediction: 
\begin{enumerate*}[label=(\roman*)]
    \item \textit{short-term}, \ie based on absolute errors of the immediate-next one hour (60\,min)
    \item \textit{intermediate-term}, \ie based on absolute errors of the immediate-next one day (1,440\,min)
    \item and \textit{long-term}, \ie based on absolute errors of the immediate-next five days (7,200\,min)
\end{enumerate*}.\footnote{\emph{Short-term} and \emph{intermediate-term} {prediction} performances evaluate {the} model usability for building control and making pricing decisions respectively. For details refer to \sref{sec:Performanceresults}.}

To analyze the accuracy of estimated flexibility metrics, we calculate the Mean Absolute Error (MAE) for the estimated $\widehat{\SoC}$ (\eref{eqn:pihat}). Additionally, we study the available and estimated operational flexibility for the given prediction horizon ($short$-$term$, $intermediate$-$term$, $long$-$term$). This is presented as a 2D region defined by $\RoC$ on the x-axis and $\SoC$ on the y-axis  \cite{Ulbig_2015_op_flex}. Available operational flexibility is defined by the operational limits of the cooling tower (\sref{sec:data}).
We compare the proposed models' evaluation time (\ie time required to predict basin temperature for each timestep) and evolution of the training loss with the number of iterations.
\section{Experimental Results}
\label{sec:results}
\subsection{Performance of grey-box models (\qref{q:greyboximprovement}--\qref{q:LSTMimprovement})}
\label{sec:Performanceresults}
\begin{figure}[!tb]
    \centering
    \includegraphics[width=0.48\textwidth]{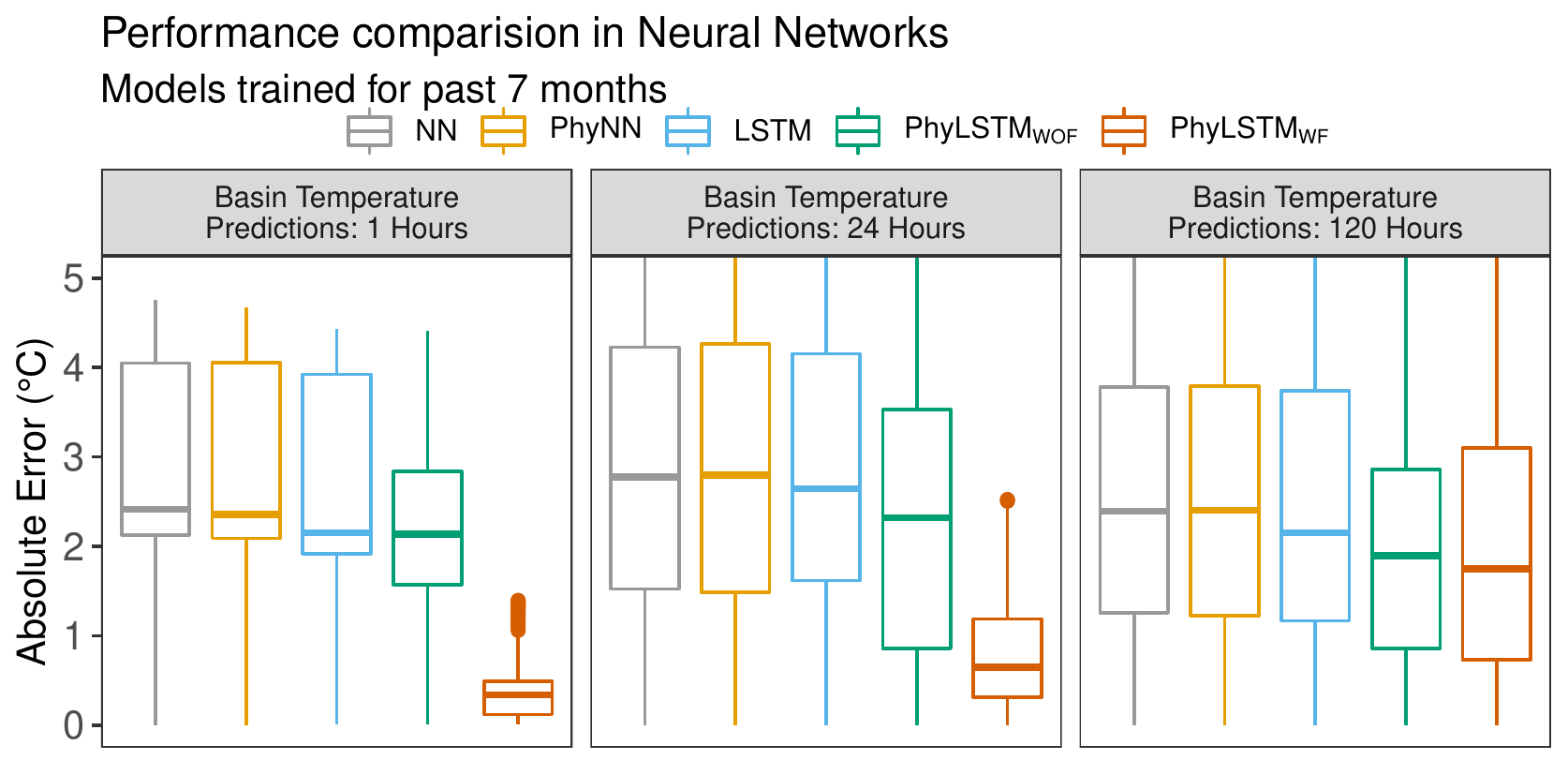}
    \caption{Absolute error for short-term, intermediate-term, and long-term models trained for seven months (210 days). Each box contains data for all test sets. (For example, a box in the long-term has: 5 test sets$\times$60 minutes$\times$120 hours.)}
    \label{fig:Results_termcompare}
\end{figure}

To answer \qref{q:greyboximprovement} and \qref{q:LSTMimprovement}, we study the absolute errors encountered in the data-driven models for all three prediction horizon (\emph{short-term}, \textit{intermediate-term}, and \textit{long-term}).
\Fref{fig:Results_termcompare} compares the absolute errors in prediction for data-driven models optimized using data from \textit{training-period} of seven months (210 days). Absolute errors for all 5 validation sets are used to construct each box.

For \textit{short-term}, black-box {\NN} and grey-box {\PyNN} have statistically similar absolute errors. Variance of grey-box {\PyLSTM$_\textit{WOF}$} (physics informed {\LSTM} without feedback of basin temperature, \sref{sec:PyLSTM}) is significantly lower compared to the black-box {\LSTM}, indicating that enforcing physics improves the precision of the model. Among all models, we note the lowest absolute errors for {\PyLSTM$_\textit{WF}$}, where the physics is enforced and feedback is applied. Since the performance of models of the \textit{short-term} horizon impacts the control effectiveness based on these models,  we can conclude that {\PyLSTM$_\textit{WF}$}~has a high potential for control design, with less than 5\% error across all validation sets. 

\emph{Intermediate-term} predictions ({the next} 24\,h) can be used to make pricing decisions based on day-ahead prices, required power, \etc
Both Physics based {\LSTM} networks  ({\PyLSTM$_\textit{WF}$} and  {\PyLSTM$_\textit{WOF}$}) outperform the black-box {\NN} and {\LSTM} {for} \textit{intermediate-term}, however, compared to \textit{short-term} the error increases.
We also note {the} error reduction {for} {\PyLSTM$_\textit{WOF}$} compared to the black-box {\LSTM}, indicating improvement in model performance by enforcing system dynamics into the network.
Similar conclusions can be drawn from the data-driven models {for} \textit{long-term}  performance, where we note statistically similar performance for both physics informed {\LSTM} networks.

While reduction in modeling error by integrating physics on neural networks has been noted in the past~\cite{raissi_2019_pinn}, our {\PyNN} and {\NN} have similar errors in all prediction periods for the selected {\ECS} configuration (2 fans).  

\subsection{Impact of training data (\qref{q:Impacttraindata})}
\label{sec:Traindataresults}
To study the impact of the amount of training data on performance, optimization, and evaluation time of data-driven models (\qref{q:Impacttraindata}), we study the absolute errors and training loss for all \textit{training periods} ($\in \{1, 3, 5, 7\}$). \fref{fig:Results_day} shows the absolute errors for \textit{intermediate-term} (24 hours) for all \textit{training periods}. Solid colored lines represent the average absolute error across all validation sets, and the grey area is the 25 to 75 percentile.

\subsubsection{{Accuracy comparison}}
\begin{figure}[!tb]
    \centering
    \includegraphics[width=0.48\textwidth]{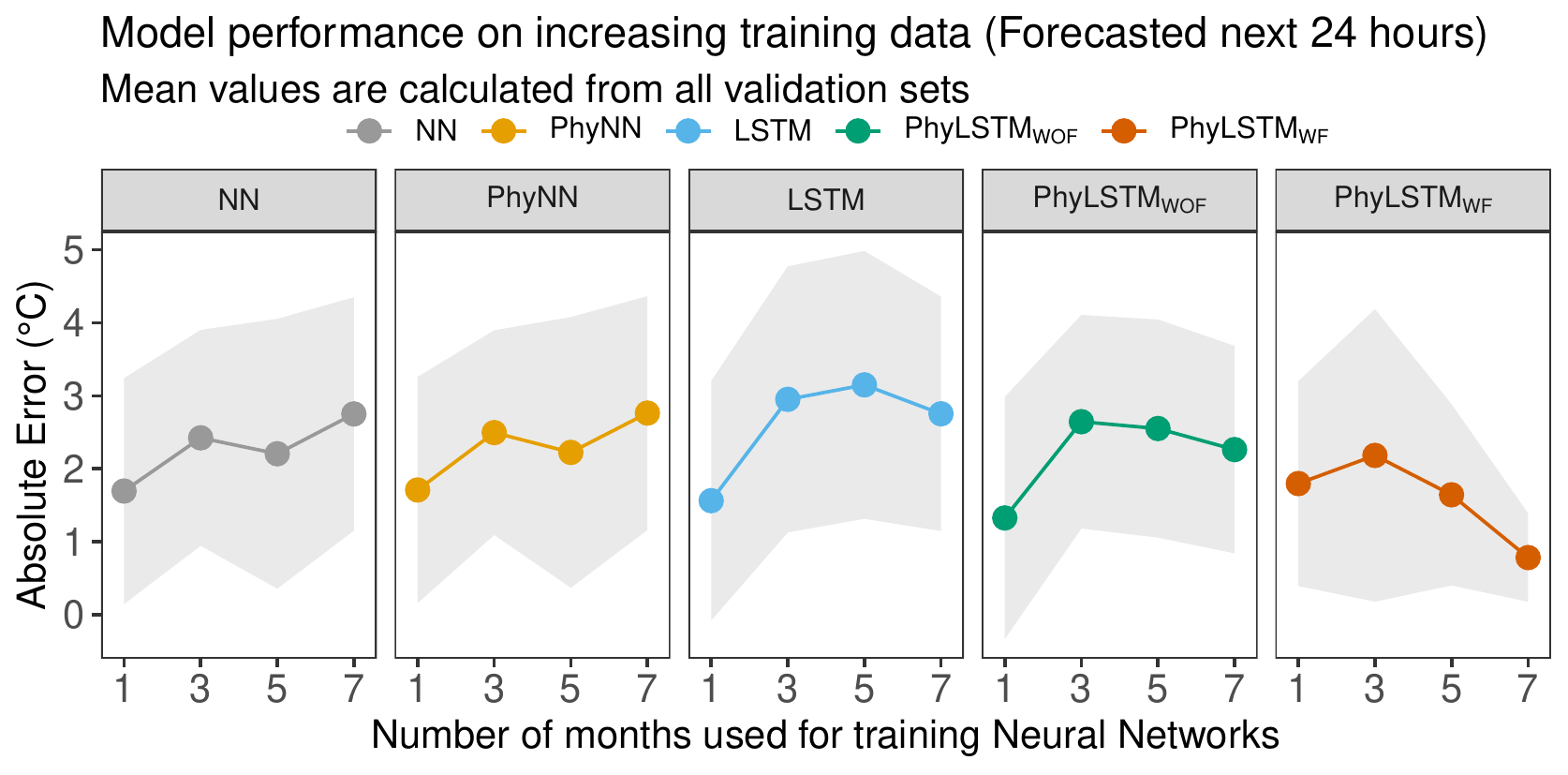}
    \caption{\small{Absolute errors for intermediate-term basin temperature predictions (24 hours) for all training periods. Each training month has 30 days of training data. Coloured line: Average absolute error; Shaded region: 25-75 percentile.}}
    \label{fig:Results_day}
\end{figure}

{We compare the accuracy of models using absolute error encountered in estimation}. Increasing the amount of training data improves the performance of the grey-box {\PyLSTM$_\textit{WF}$} model, \ie we get the lowest absolute errors for {\PyLSTM$_\textit{WF}$} trained with seven months of training data. The performance of the {\PyLSTM$_\textit{WF}$} improves with the size of training data as seen in~\fref{fig:Results_day} (decrease in average and variance of error across all validation sets). This happens because the model learns the time-dependent behavior of the process accurately from the system dynamics and the network feedback. We notice a similar performance characteristics for \textit{short-term} and \textit{long-term}. 

The absolute error for black-box {\NN} and grey-box {\PyNN} is higher for models trained for 7 months of training data, in contrast to models trained for 1 month of training data. The data representing system dynamics of immediately preceding 1 month train a better model compared to the immediately preceding 7 months, because for 7 months the model has to generalize for multiple months (\ie 7 months include various seasons, that have different outside temperature ranges). Their performance for training data for 3 and 5 months remains statistically similar.\footnote{Based on Wilcoxon test for testing statistical significant difference} Additionally, we note no significant impact of training data on the variance of errors in both {\NN} and {\PyNN}.

\subsubsection{{Training loss comparison}}
We study the evolution of training loss during training to see how different neural networks are optimized. 
In the case of a black-box {\NN}, {the} loss consists of a mean square error {\MSE} on the temperature of the basin (\eref{eqn:NN_loss}). Training loss converges in approximately 500 iterations with 1 month of training data. Additionally, it takes fewer iterations for loss to converge for more training data, \ie 3, 5, 7 months.

{The l}oss for grey-box models ({\PyNN} and  {\PyLSTM$_\textit{WF}$}) comprises the mean squared error in {\BasinTemp} and \textit{f} (\eref{eqn:PyNN_loss} and \eref{eqn:PyLSTM_loss}).
We notice that  enforcing the dynamics of the process in the neural network models provides faster convergence, \ie training loss converges in approximately 300 iterations for {\PyNN} and 150 iterations for  {\PyLSTM$_\textit{WF}$}{,} with 1 month of training data (compared to 500 iterations {for} {\NN}).
Furthermore, {\PyLSTM$_\textit{WF}$} also converges in fewer iterations with more training data.
It takes approximately 100 iterations for the training loss to converge with a training data window of 7 months.
We can conclude that  {\PyLSTM$_\textit{WF}$} converges faster with more training data in contrast to {\PyNN} and {\NN}.

\subsubsection{{Computation time comparison}}
We compare the {computation} time of data-driven models.
Training time is higher in the case of  {\PyLSTM$_\textit{WF}$} when compared to {\NN} and {\PyNN}. One reason for this is the higher complexity of {\LSTM} cells in  {\PyLSTM$_\textit{WF}$} compared to the simple activation-based neurons in the other two models. This complexity results in more calculations while training the {\LSTM} based models, increasing the overall training time. Each iteration for {\PyLSTM$_\textit{WF}$} takes approximately 0.15\,s compared to 0.02\,s in case of {\PyNN} and 0.001\,s in case of {\NN}.\footnote{Training and evaluation are done with systems using {an} Intel Xeon E5645 and a single 4 core E3-1220v3 (3.1GHz) with 16\,GB {of} RAM.}
For our models, we notice that {\NN} and {\PyNN} have the shortest evaluation time, with less than 5\,ms on average to make each prediction. 
{For} {\PyLSTM$_\textit{WF}$} {it} takes 15\,ms on average to make each prediction. For example, it takes 0.09\,s (=~60 $\times$ 15\,ms) for \textit{short-term} (60 timeslots).
We noted that {for all proposed models} the time to generate each prediction is small ($\ll$ 1\,s) compared to the one-minute timeslot. 


In summary, integrating physics provides faster convergence in data-driven grey-box models compared to the black-box {\NN}. In grey-box models,  {\PyLSTM$_\textit{WF}$} has a lower prediction error compared to all other models, and overall less than 5\% error compared to the white-box model. With more training data, the accuracy of  {\PyLSTM$_\textit{WF}$} increases, and it reaches convergence earlier.
\begin{figure}[tb!]
    \centering
    \includegraphics[width=0.48\textwidth]{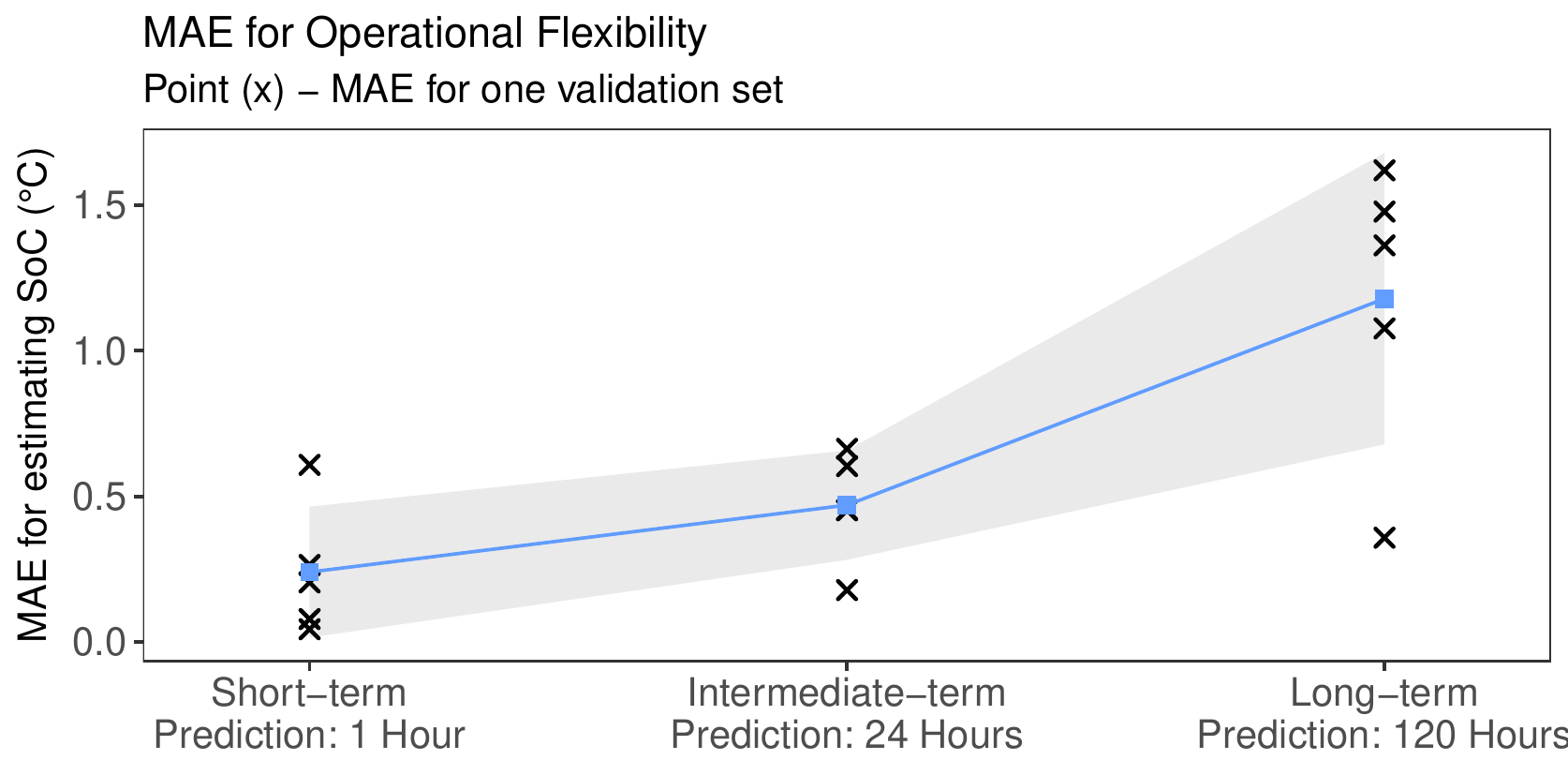}
    \caption{MAE in State of Charge (${\SoC}$) for {\PyLSTM$_\textit{WF}$}.}
    \label{fig:OPFlex_pi_MAE}
\end{figure}
\begin{figure}[tb!]
    \centering
    \includegraphics[width=0.45\textwidth]{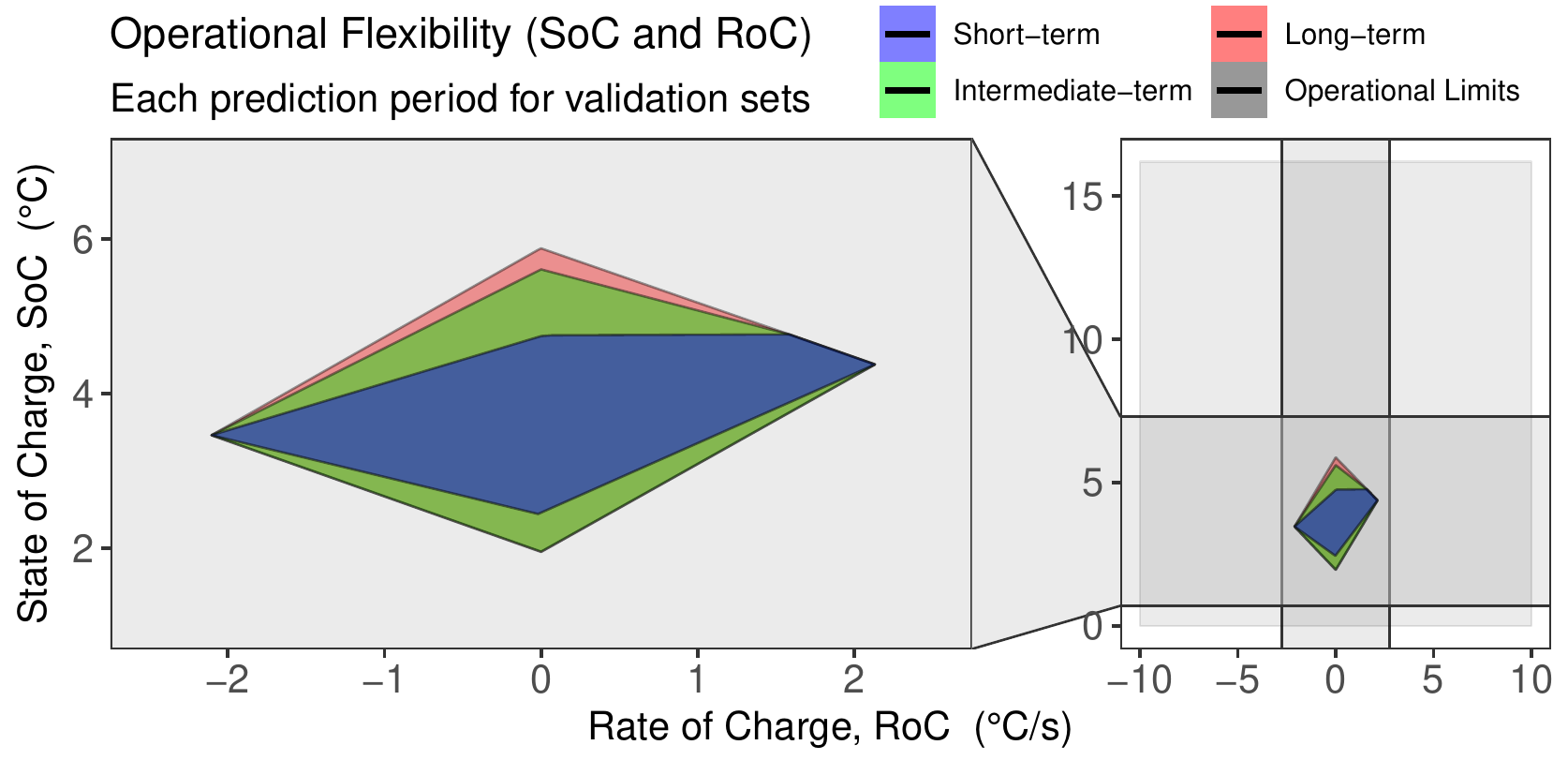}
    \caption{\small{Available and estimated operational flexibility for {\PyLSTM$_\textit{WF}$}. (Operational limits are defined by upper and lower limits of {$\SoC$ and {$\RoC$}}) }}
    \label{fig:OPFlex_area}
\end{figure}
\vspace{-2mm}
\subsection{Operational flexibility (\qref{q:opflex})}
\label{sec:opflexresults}
We calculate operational flexibility metrics ${\SoC}$ and ${\RoC}$ using the estimates from our data-driven  {\PyLSTM$_\textit{WF}$} to answer (\qref{q:opflex}). \Fref{fig:OPFlex_pi_MAE} provides the Mean Absolute Errors (MAE) encountered in estimating $\widehat{\SoC}$ for different prediction periods. These results are for the {\PyLSTM$_\textit{WF}$} trained with 7 months of training data. The MAE is lower in \textit{short-term} predictions compared to \textit{intermediate-term} and \textit{long-term} predictions. 

We present the estimated and available operational flexibility in~\fref{fig:OPFlex_area}, where the $x$-axis is ${\RoC}$ and the $y$-axis is ${\SoC}$. The grey area represents operational limits that define the available operational flexibility (\sref{sec:data}). White-box model and  {\PyLSTM$_\textit{WF}$} provide similar regions for estimated operational flexibility. Additionally, we notice that the area of this region (representing estimated flexibility) increases with the period of prediction, \ie larger flexibility in the \textit{long-term} compared to the \textit{short-term}. {This} behavior {is due to} the longer duration, which allows for a {broader} range of operational settings during data simulation.

\section{Conclusions}
\label{sec:conclusions}
Identifying system behavior in an energy-intensive Evaporative Cooling System ({\ECS}) is critical for flexibility identification, building control policies, \etc In this paper, we define novel data-driven models for system identification in {\ECS}. More specifically 
\begin{enumerate*}[label=(\roman*)]
    \item  we define grey-box physics informed networks (\PyNN, \PyLSTM$_\textit{WOF}$ and \PyLSTM$_\textit{WF}$)
    \item we choose an industrial process, \ie an induced draft cooling tower, that offers operational flexibility to define a white-box model to simulate training data, and
    \item  we evaluate the performance of our proposed grey-box models by training them for different training data sizes and validation sets.
\end{enumerate*}
From our analyses, we conclude the following:

\begin{enumerate}[label={(\arabic*)}]
    \item  {\PyLSTM$_\textit{WF}$}, \ie a physics informed {\LSTM} model with feedback (\sref{sec:PyLSTM}) outperforms the grey-box ({\PyNN} and {\PyLSTM$_\textit{WOF}$}) and black-box ({\NN} and {\LSTM}) models, where we see that it has less than 2\% error {for a \emph{short-term}} prediction horizon (\fref{fig:Results_termcompare}). Similar performance is observed {for an \emph{intermediate-term}} prediction horizon, and we can conclude that {a} physics informed {\LSTM} has the highest accuracy among the data-driven models.
    \item The performance of the {\PyLSTM$_\textit{WF}$} improves with longer training data window{,} as opposed to {\PyNN} and {\NN} (as noted by {the} decrease in average and variance of absolute error in~\fref{fig:Results_day}). Furthermore, {\PyLSTM$_\textit{WF}$} {can be trained} in fewer iterations with more data. It takes approximately 100 iterations for {the} training loss to converge {for} 7 months of training data.
    \item The computation time to predict the future state of the process is insignificant ($\ll$ 1\,s) compared to the time it takes to evaluate and implement control decisions {in demand response (order of minutes)}. Such property makes the grey-box models suitable {to be used to develop} the control policies.
\end{enumerate} 

Overall the proposed physics informed {\LSTM} network has excellent potential for developing flexibility identification that can be employed for estimating operational flexibility in an industrial energy intensive process such as {an} {\ECS} (\fref{fig:OPFlex_area}).
In future research, we will
\begin{enumerate*}[label=(\roman*)]
    \item evaluate these grey-box models' potential in developing effective control policies for industrial systems,
    \item explore these networks' performance for multiple inputs and multiple outputs systems compared to the single output system that we tested, and
    \item {quantitatively assess the effectiveness of PhyLSTM$_\textit{WF}$ in other real-world applications}
\end{enumerate*}.

\section*{Acknowledgment}
\normalsize
Part of the research leading to these results has received funding from Agentschap Innoveren $\&$ Ondernemen (VLAIO) as part of the Strategic Basic Research (SBO) program under the InduFlexControl project and the European Union's Horizon 2020 research and innovation program for the project BIGG (grant agreement no.\ 957047)

\bibliographystyle{IEEEtranN}
\footnotesize
\bibliography{references}


\begin{IEEEbiography}[{\includegraphics[width=1in,height=1.5in,clip,keepaspectratio]{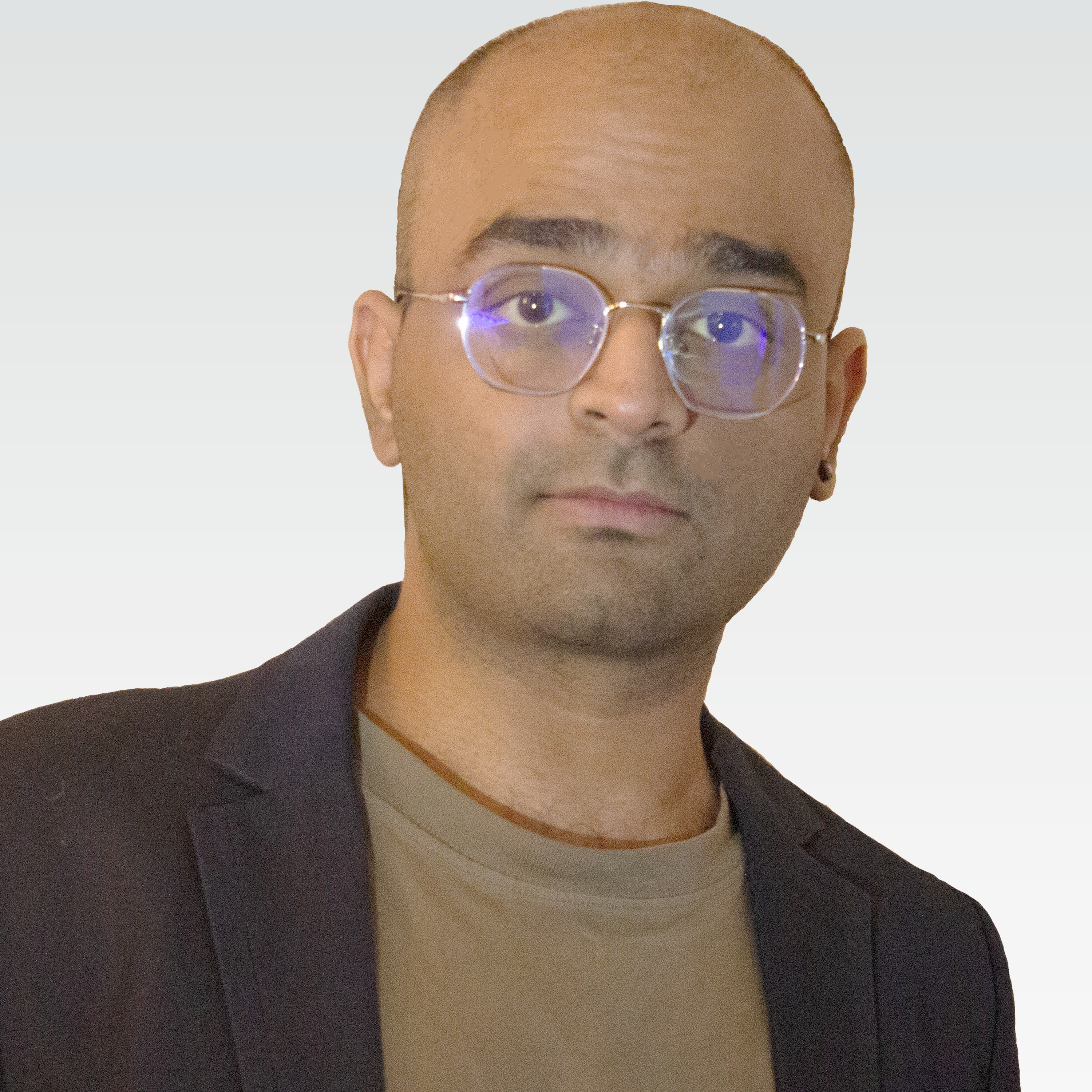}}]%
{Manu Lahariya} is working towards a Ph.D. degree in computer science engineering (specializing in machine learning for energy applications) with the research group Artificial Intelligence for Energy (AI4E) at IDLab, Ghent University, Ghent, Belgium. He received M.Tech. degree and B.Tech. degree in aerospace engineering from the Indian Institute of Technology, Kharagpur, India in 2016 and 2017, respectively. He has contributed to multiple European research projects (e.g., BIGG, InduFlex) and was a visiting researcher at Robust Autonomy and Decisions Lab, University of Edinburgh, Edinburgh, Scotland. His research interests include physics-based machine learning, reinforcement learning, deep learning, and statistical modeling for designing efficient control and system identification across different applications (e.g., smart grids, residential space heating, soft robotics, etc.).
\end{IEEEbiography}

\begin{IEEEbiography}[{\includegraphics[width=1in,height=1.5in,clip,keepaspectratio]{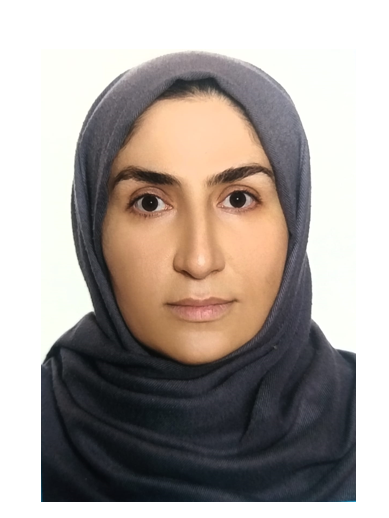}}]%
{Farzaneh Karami} received the MSc degree in Control Engineering from the Iran University of Science and Technology, Tehran, Iran, in 2010, and pursued her PhD degree in Computer Science (specializing in Optimization and Operations Research) at KU Leuven, Belgium. Since 2020, she has been a postdoctoral researcher with the Department of Electromechanical, Systems and Metal Engineering, Ghent University, Core lab EEDT-DC, Flanders Make, Belgium. Her research interests include using tools from mathematical optimization to create natural, understandable, and data-driven solutions that provide essential insights into improving operations.
\end{IEEEbiography}

\begin{IEEEbiography}[{\includegraphics[width=1in,height=1.5in,clip,keepaspectratio]{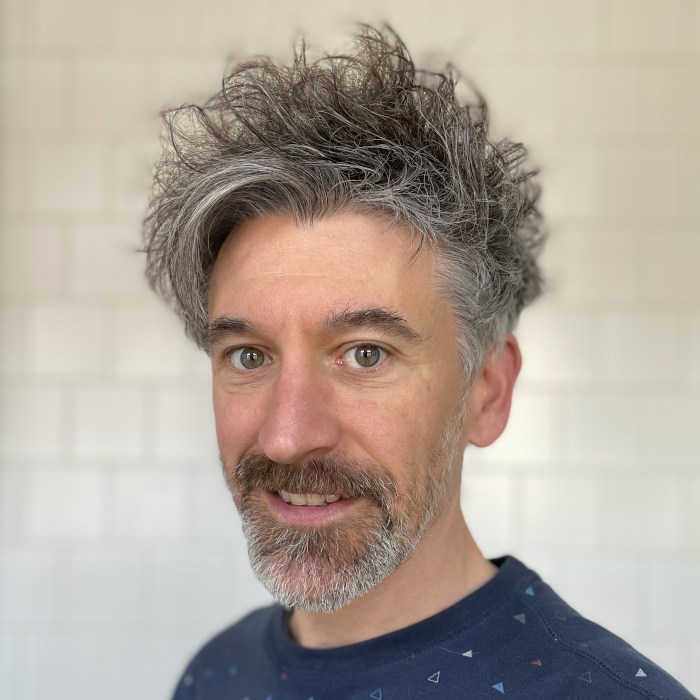}}]%
{Chris Develder} is associate professor with the research group IDLab in the Dept. of Information Technology (INTEC) at Ghent University – imec, Ghent, Belgium. He received the MSc degree in computer science engineering and a PhD in electrical engineering from Ghent University (Ghent, Belgium), in Jul. 1999 and Dec. 2003 respectively (as a fellow of the Research Foundation, FWO). He has stayed as a research visitor at UC Davis, CA, USA  (Jul.-Oct. 2007) and at Columbia University, NY, USA (Jan. 2013 – Jun. 2015). He was and is involved in various national and European research projects (e.g., FP7 Increase, FP7 C-DAX, H2020 CPN, H2020 Bright, H2020 BIGG, H2020 RENergetic, H2020 BD4NRG).
Chris currently leads two research teams within IDLab, (i) UGent-T2K on converting text to knowledge (i.e., NLP, mostly information extraction using machine learning), and (ii) UGent-AI4E on artificial intelligence for energy applications (e.g., smart grid). He has co-authored over 200 refereed publications in international conferences and journals. He is Senior Member of IEEE, Senior Member of ACM, and Member of
ACL.
\end{IEEEbiography}

\begin{IEEEbiography}[{\includegraphics[width=1in,height=1.5in,clip,keepaspectratio]{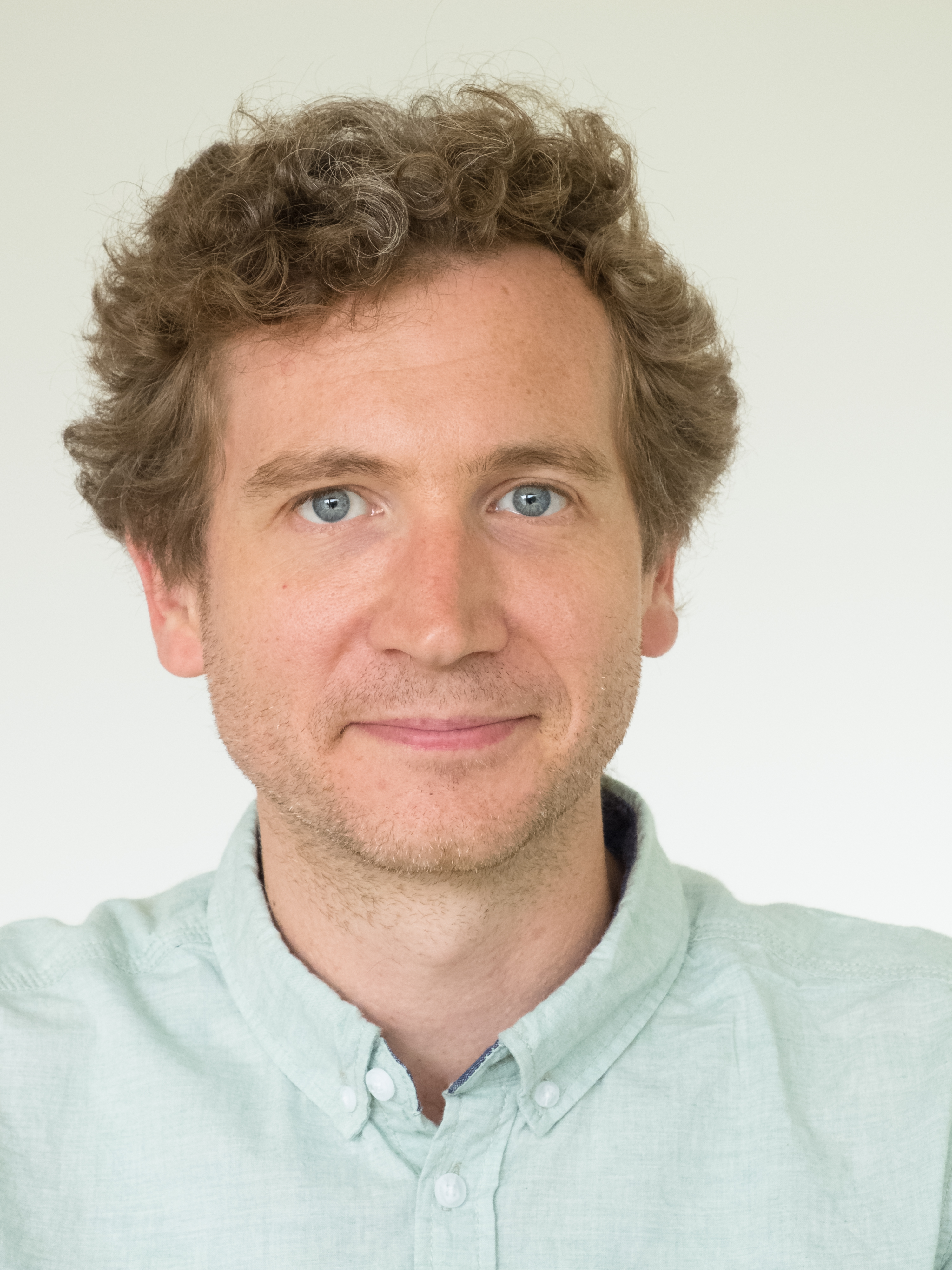}}]%
{Guillaume Crevecoeur} (°1981) received his Master and PhD degree in Engineering Physics from Ghent University in 2004 and 2009, respectively. He received a Research Foundation Flanders postdoctoral fellowship in 2009 and was appointed Associate Professor at Ghent University in 2014. He is member of Flanders Make in which he leads the Ghent University activities on sensing, monitoring, control and decision-making. With his team, he conducts research at the intersection of system identification, control and machine learning for mechatronic and industrial robotic systems. His goal is to endow physical dynamic systems with improved functionalities and capabilities when interacting with uncertain environments, other systems and humans. 
\end{IEEEbiography}

\end{document}